\documentclass[namedreferences,hyperref,optionalrh,solaromanenum]{spr-sola}

\usepackage{graphicx}                    
\usepackage{amssymb}                    
\usepackage{color}                       

\usepackage{xcolor}
\usepackage{todonotes}

\chardef\us=`\_

\begin{document}

\begin{frontmatter}

\title{Investigation of the Two-Dimensional Velocity Field of the Large-Scale Coronal Wave from September
6, 2011 using the SOLERwave Tool}

%

\author[corref,addressref={aff1},email={m.baumgartner-steinleitner@uni-graz.at}]{\inits{M. }\fnm{Markus } \snm{Baumgartner-Steinleitner}\orcid{0009-0004-2379-8067}}

\author[addressref={aff1,aff2},email={astrid.veronig@uni-graz.at}]{\inits{A. }\fnm{Astrid M. } \snm{Veronig}\orcid{0000-0003-2073-002X}}

\author[addressref={aff1},email={karin.dissauer@uni-graz.at}]{\inits{K. }\fnm{Karin } \snm{Dissauer}\orcid{0000-0001-5661-9759}}

\author[addressref={aff3},email={jens.pomoell@helsinki.fi}]{\inits{J. }\fnm{Jens } \snm{Pomoell}\orcid{0000-0003-1175-7124}}

%
\runningauthor{M. Baumgartner-Steinleitner et al.}
\runningtitle{SOLERwave: 2D Velocity Field of Coronal Waves}

\address[id=aff1]{Institute of Physics, University of Graz, Universitätsplatz 5, Graz,
8010, Austria}
\address[id=aff2]{Kanzelhöhe Observatory for Solar and Environmental Research, University of Graz, Austria}
\address[id=aff3]{Department of Physics, University of Helsinki, Finland}

\begin{abstract}


We investigate the two-dimensional velocity field of the fast and complex large-scale coronal wave observed on September 6, 2011. We use both a classical perturbation profile approach and the newly developed multi-sector method of the SOLERwave tool, using a Huygens-plotting-based approach.  The multi-sector method utilizes perturbation profiles derived in multiple directions (sectors) to determine the location of the wavefront at a given time. The two-dimensional velocity vector at each point along the wavefront is derived by identifying the point closest to it along the wavefront observed one time step earlier and dividing the distance between the two points along the solar surface by the time difference between the observations. For the event under study the resulting two-dimensional velocity field shows a significant difference between the northward traveling and the northwest ward traveling part of the wave front of over 40\%, in the range of 750 to 1500 km/s. To determine the cause of this difference in speed, we investigate the coronal structures, the photospheric magnetic field distribution and the Alfv\'{e}n speed derived from a steady-state coronal magneto hydro dynamic (MHD) solution along different propagation directions of the wave. We find the large difference in magnetosonic speed found in the investigated sector as the most likely explanation for the velocity difference.

\end{abstract}

%

\end{frontmatter}

\section{Introduction}

Large-scale coronal waves have been observed for nearly 30 years. From their first observations by \citet{moses_eit_1997} and \citet{thompson_sohoeit_1998} using the Extreme-Ultraviolet Imaging Telescope (EIT; \citealt{delaboudiniere_eit_1995}) onboard the Solar and Heliospheric Observatory
(SOHO; \citealt{domingo_soho_1995}) to more recent observations using the Extreme Ultraviolet Imager (EUVI; \citealt{howard_sun_2008}) onboard the Solar Terrestrial Relations Observatory (STEREO; \citealt{kaiser_stereo_2008}) and the Atmospheric Imaging Array  (AIA; \citealt{lemen_atmospheric_2012}) onboard the Solar Dynamics Observatory (SDO; \citealt{pesnell_solar_2012}), hundreds of waves have been investigated, either in detailed case studies (e.g. \citealt{vrsnak_multi-wavelength_2006,veronig_high-cadence_2008,kienreich_stereo_2009,ma_new_2009,veronig_first_2010, vanninathan_coronal_2015, dissauer_projection_2016,devi_extreme-ultraviolet_2022,dresing_17_2023,mann_propagation_2023}), or in statistical investigations (e.g. \citealt{warmuth_kinematical_2011, nitta_large-scale_2013, muhr_statistical_2014}). The current conclusion from these analyses on the nature of large-scale coronal waves is that they are fast, large-amplitude or shocked fast-mode magnetosonic waves traveling in the corona of the Sun (see reviews by \citealt{zhukov_eit_2011,patsourakos_nature_2012,warmuth_large-scale_2015, long_statistical_2017}). Various case studies have shown that the fast lateral expansion of the flanks of the associated coronal mass ejection (CME) act as exciting agent \citep{kienreich_stereo_2009, patsourakos_extreme_2009, veronig_genesis_2018}. 

Large-scale coronal waves are often linked with Moreton waves originally discovered by \citet{moreton_recent_1960}, which are interpreted as the imprint on the chromosphere from the pressure pulse exerted by the coronal wave, that compresses and pushes the chromospheric plasma downward \citep{uchida_propagation_1968,vrsnak_flare_2002,warmuth_large-scale_2015}. While some large-scale coronal waves are observed as quasi-circular, most exhibit a non-isotropic propagation \citep{thompson_catalog_2009,liu_advances_2014,warmuth_large-scale_2015,long_localized_2021}. As expected for magnetosonic (shock) waves, large-scale coronal waves exhibit ``wave like'' behavior like refraction, reflection and transmission in regions with a different fast mode speed such as active regions and coronal holes, due to different magnetic flux density and plasma density \citep{gopalswamy_euv_2009,olmedo_secondary_2012,piantschitsch_numerical_2017,liu_impacts_2018,zhou_total_2022}. Further, they have been observed interacting with features in the corona like coronal loops, filaments and streamers, increasing their intensity or exciting an oscillation \citep{vanninathan_coronal_2015, warmuth_large-scale_2015, liu_advances_2014}.


Large-scale coronal waves have been mainly investigated either by visually tracking the wave front (e.g. \citealt{warmuth_multiwavelength_2004,veronig_high-cadence_2008,thompson_catalog_2009}) or algorithmically using the so-called ring method or perturbation profile method (e.g. \citealt{warmuth_evolution_2001,podladchikova_automated_2005, muhr_analysis_2011, nitta_large-scale_2013, long_corpita_2014, vanninathan_coronal_2015, dissauer_projection_2016,dresing_17_2023}), or derivatives there of. In essence, the perturbation profile method uses the increase in intensity due to the passing wavefront, measured along great arcs from a presumed wave origin. To enhance transient features in the images, often base ratio or base difference images are used. To improve pixel statistics and reduce noise, the perturbations are calculated as the average value within segments of equal angular length and width along the great arc. Waves manifest in these perturbation profiles as local maxima that are propagating in time. For further analysis, these peaks are sometimes fitted with a Gaussian function \citep{muhr_analysis_2011, dissauer_projection_2016, long_corpita_2014,liu_advances_2014,pesce-rollins_coupling_2022}. From these fitted profiles, the amplitude of the wave, its width as well as its distance can be derived, and consequently also its  speed (e.g. \citealt{veronig_first_2010,muhr_analysis_2011,dissauer_projection_2016}).  Compared to the visual tracking of coronal wave fronts, the algorithmic nature of the perturbation profile method allows objective, reproducible and automatic calculation of the main characteristica of the wave.

Up to now, a few algorithmic approaches have been developed to detect coronal wave fronts and to study their characteristics. The NEMO algorithm developed by \citet{podladchikova_automated_2005} divides the solar sphere into 8 sectors, centered around the flare position, in which the perturbation profiles are calculated to determine the fastest radial speed of the wave front. Solar Demon \citep{kraaikamp_solar_2015} and \citet{nitta_large-scale_2013} use 24 non overlapping sectors for the perturbation profile calculation to determine the radial speed. CORPITA by \citet{long_corpita_2014} uses perturbation profiles of overlapping sectors to determine the radial wave speed in a fine radial grid of $1$° resolution. AWARE by \citet{ireland_aware_2019} finds the wave front on a cartesian grid using persistence ratio images, but reduces the information to a single mean radial speed of the wave front. Soler Demon, CORPITA, AWARE and NEMO focus on the automatic detection of large-scale coronal waves rather than the determination of the velocity vector field of the front. A recent method by \citet{rigney_tracking_2024} uses the Wavetrack algorithm by \citet{stepanyuk_multi-scale_2022} to determine the position and extent of the wave combined with a Fourier local correlation tracking to estimate the two dimensional velocity field of the full wave front between two snapshots, but reduces this information to a single value using a mass velocity calculation. To the knowledge of the authors, the only studies that calculated a full velocity field using the Huygens plotting approach were done in  \citet{wills-davey_observations_1999} and \citet{wills-davey_tracking_2006}, but it has been used only on a very limited set of observations.

In this paper, we present the newly developed SOLERwave tool, consisting of a single-sector analysis of perturbation profiles with integrated wave tracking and a multi-sector analysis. The multi-sector analysis combines multiple perturbation profiles into one observation to determine the velocity vector field of the wave front using a version of Huygens plotting \citep{wills-davey_tracking_2006}.  We apply this method to analyze the complex and fast large-scale coronal wave observed at 22:20~UT on the September 6, 2011, to in particular investigate the velocity field of the coronal wave in different directions.

The paper is structured as followed. In Section~\ref{sec:data}, we discuss the data used in the case study as well as the data preprocessing. In Section~\ref{sec:methodes}, the single-sector and multi-sector method of the SOLERwave tool are described. Section~\ref{sec:results} shows the results of the case study, which are discussed in Section~\ref{sec:discussion}. In the Appendix, we present an in-depth discussion on the various uncertainty terms that affect of the wave kinematics.

\section{Data and Processing}
\label{sec:data}

The large-scale coronal wave that occurred on the September 6, 2011 is associated with a X2.1 class (GOES (Geostationary Operational Environmental Satellites) flares start time: 22:12 UT) event in the NOAA active region 11283 and a CME with a speed of $990$~km~s$^{-1}$ \citep{dissauer_projection_2016}. The event was observed in quasi quadrature by SDO and STEREO-A, which were separated by about $103$° in longitude on that day.

Large-scale coronal waves can be observed in multiple wavelengths with the best signal observed in extreme ultra violet (EUV) filters that image the solar corona at temperatures around $1$ - $2$ MK. In this study, we use the SDO/AIA $211$~\AA~passband, which is dominated by emission from Fe XIV with a formation temperature of about 2 MK \citep{lemen_atmospheric_2012}. As the large-scale coronal wave temporarily vanished in the AIA $211$~\AA~passband (see \citealt{dissauer_projection_2016} for a detailed analysis), complementary data from the SDO/AIA $335$~\AA~passband (Fe XVI, $T \sim 2.5$ MK) is used, in which the wave does not vanish. For both channels, only images with an exposure time greater than $1.5$ s were used due to their higher signal to noise ratio, resulting in an effective observation cadence of $24$ s during the wave event.

As a first step of the preprocessing, all images are normalized by dividing by their exposure time. The images are then corrected for solar differential rotation with SunPy's \citep{sunpy-community_sunpy_2023} \textit{propagate\_with\_solar\_surface} function using the model by \citet{howard_solar_1990} with respect to the pre-event image and then rebinned by $2\times2$ from $4096 \times 4096$ pixels to $2048 \times 2048$ pixels. In the last step of preprocessing, the base ratio images are created by dividing each frame of the sequence by a pre-event image. For the single-sector analysis in AIA $211$~\AA, the image at 22:15:50~UT is chosen, as it marks the start of the impulsive phase of the flare \citep{dissauer_projection_2016}. Minimizing the time between the base image and frame under investigation reduces the changes due to fluctuations in the corona not related to the wave in the base ratio images, improving observation. For the multi-sector analysis, a later base image (22:19:38~UT for AIA $211$~\AA) was chosen despite being after the flare onset, as it minimizes the effect of diffraction patterns and blooming of the CCD away from the flaring region caused by high intensity of the flare in the images. The base image for the AIA $335$~\AA~passband was chosen around the same time (22:19:53~UT).

\section{Methods: the SOLERwave Tool}
\label{sec:methodes}

To analyze the large-scale coronal wave that occurred on September 6, 2011 we use the newly developed SOLERwave tool. It allows the automatic loading, preprocessing and analysis of the wave to determine the speed, amplitude and direction of a wave front. It is split into two major components: the single-sector analysis and the multi-sector analysis.  The single-sector analysis derives the perturbation profiles for an individual sector with a chosen width and direction, identifies peaks in the profile and uses a wave tracing algorithm to track moving peaks (i.e. large-scale coronal waves) in order to derive the wave front distances as function of time and the wave speed. The multi-sector analysis applies the single-sector analysis on multiple, overlapping sectors, but uses a Huygens-plotting-based algorithm instead of the wave tracer of the single-sector method to derive the 2D velocity vectors of the wave front.

\subsection{Single-Sector Analysis}

The single-sector analysis consists of three major steps:  creating perturbation profiles,  peak finding and fitting, and wave tracing. These steps are explained in more detail below.

\subsubsection{Creating Perturbation Profiles}
\label{sec:Creating Perturbation Profiles}

\begin{figure}
   \centering
    \includegraphics[height = 5.5cm]{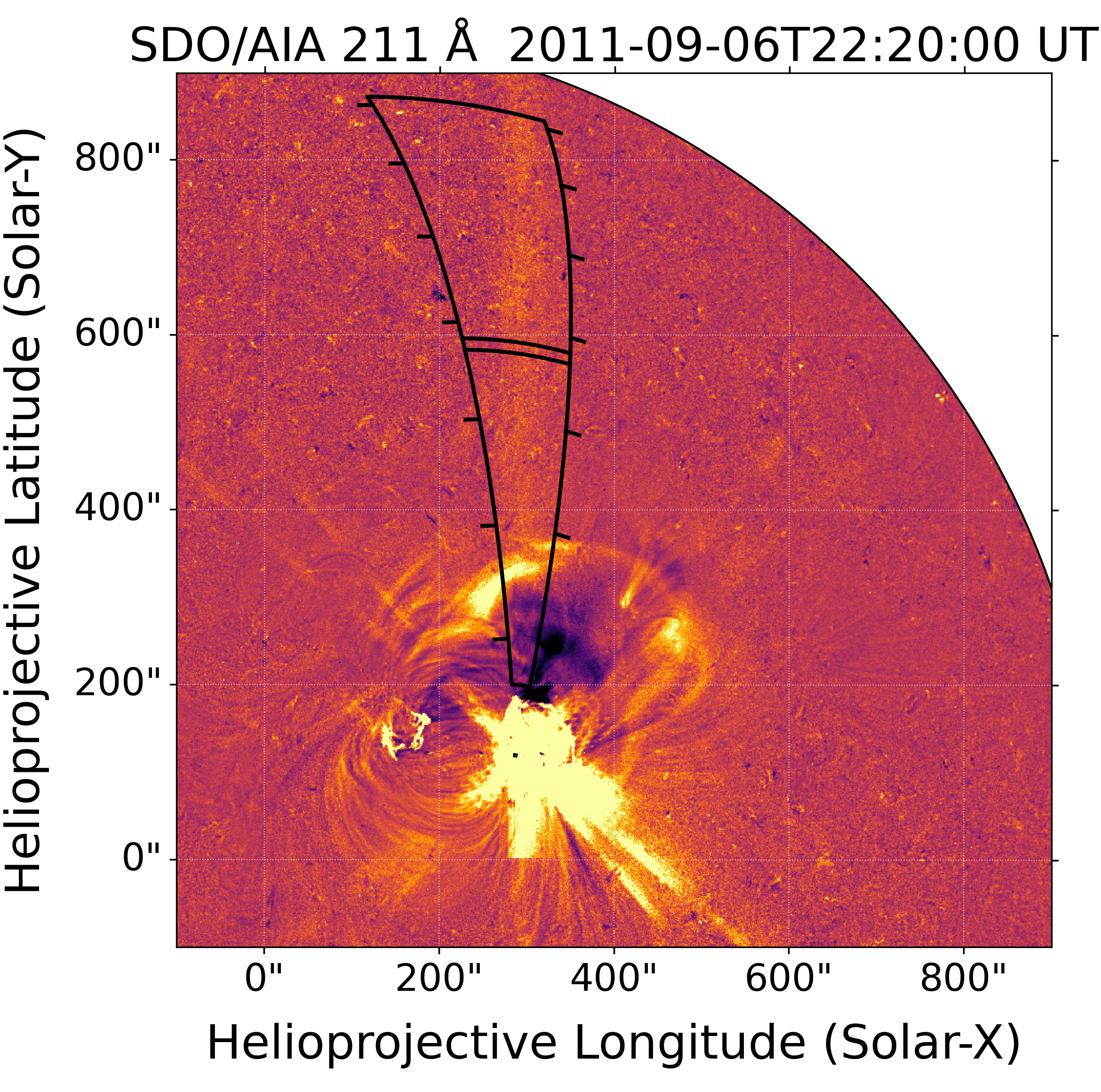}
    \includegraphics[height = 5.5cm]{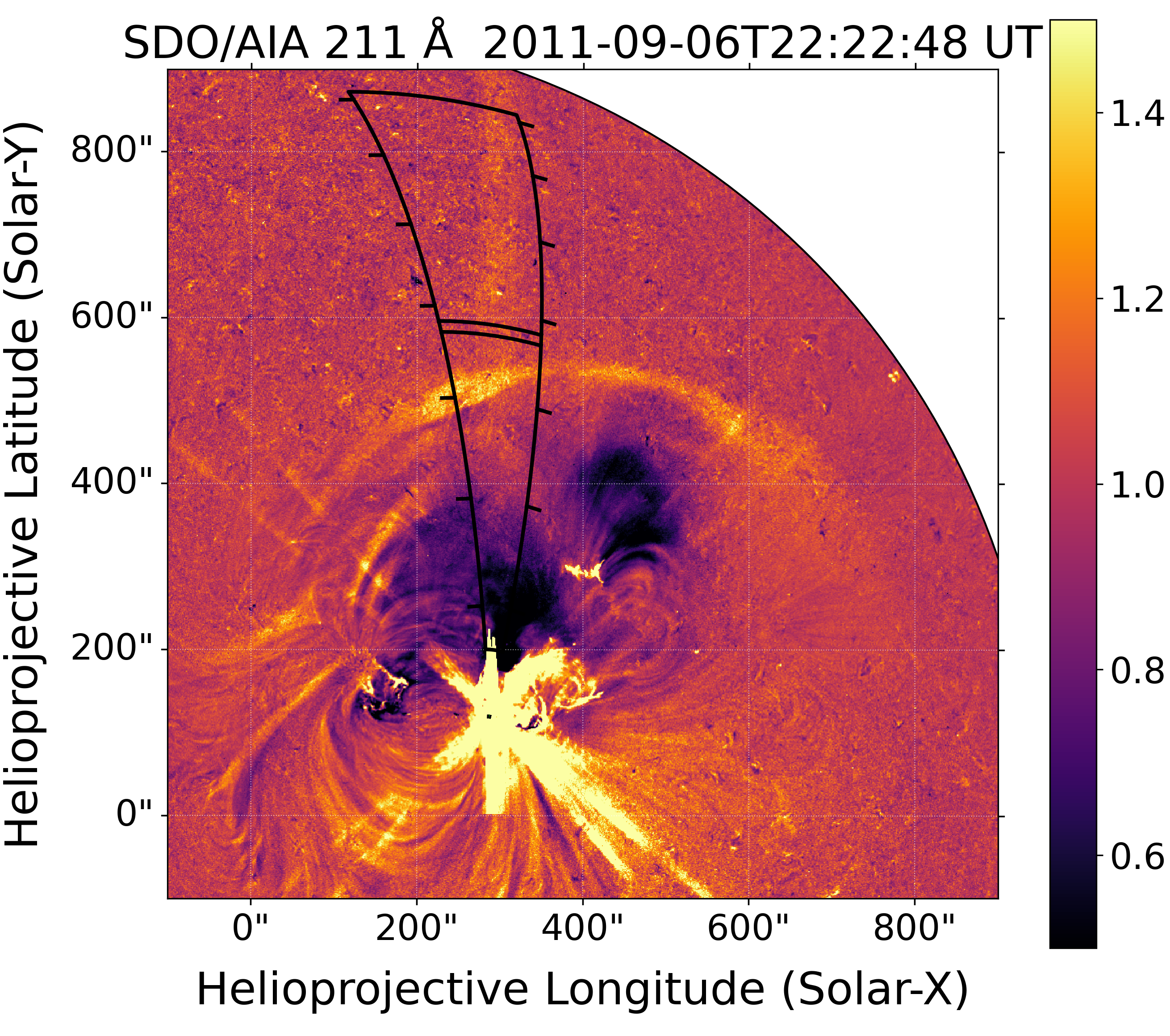}
      \caption{Two AIA $211$~\AA~base ratio images (base image at 22:15:50 UT) of the large-scale coronal wave observed on September 6, 2011. A single-sector is overplotted with an angular extent between $-15$° and $0$° with respect to the great circle intersection both the presumed wave origin at (N14°, W18°) and the solar north pole. The markers on the sector show $100$~Mm increments. Within the sector a single segment with $1$° length is marked. The wave is visible as bright circular front in the beginning, deviating toward the solar east as it travels. The online version shows the according movie over the time range 22:16 to 22:30 UT.}
         \label{fig:movie_north_wave}
\end{figure}

In order to analyze the signatures of large-scale coronal waves in the lower solar atmosphere, the distance of the wave front from its origin has to be extracted for each time step. To do so, a spherical grid is defined on the solar surface with the wave origin at its center. Following previous studies \citep{muhr_analysis_2011,nitta_large-scale_2013, vanninathan_coronal_2015, dissauer_projection_2016}, we use the position of the associated flare as an initial estimate of the wave center. Since large-scale coronal waves usually reveal a non-isotropic propagation, sectors are defined on this spherical grid given by their central direction and their width (i.e. they are constrained by an azimuthal angle). The azimuthal angle is defined between the great circle intersecting both the wave origin and the solar North. Angles are counted counter-clockwise. These sectors are further split into segments along concentric circles starting from the wave origin. To achieve sufficient pixel statistics, the sector width was chosen to be $15$°, and the segments have a length of $1$° (corresponding to about $12.1$~Mm on the solar surface). Figure~\ref{fig:movie_north_wave} shows as an example a single sector between $-15$° and $0$°, where we indicate the width of splitting by showing one of the segments.

SOLERwave divides the solar sphere into these segments considering projection effects, since what is actually observed is a projected 2D image of the 3D spherical surface of the Sun. Internally, SOLERwave uses the coordinate information in the header of the FITS images transformed to heliocentric coordinates using the AstroPy WCS (World Coordinate System) module \citep{astropy_collaboration_astropy_2022}, as well as derivative functions of it from the SunPy coordinates module \citep{sunpy-community_sunpy_2023}. Based on this information, a mask is generated which is used to perform operations on pixels of each segment across all images in the analyzed time sequence. 

The spherical masks generated are used on preprocessed data to calculate the median of the intensity for each segment. Plotting the median of all segments along one great circle results in perturbation profiles. The median was chosen as numerical tests showed it gives a better and more robust representation of the average value in case of ratio images (see Appendix Section~\ref{appendix:sec_base_ratio} for more details). To increase the number of data points along a great arc and to reduce the error on the distance caused by the $1$° step between segments, in each sector the mask is calculated also for three additional (overlapping) realizations where each realization is shifted by $0.25$° along the great arc originating from the presumed wave center. This results in a staggered output similar to a running average along the great arc direction.

\subsubsection{Peak Finding and Fitting}
\label{sec:Peak finding and fitting}

In order to trace the crest of a large-scale coronal wave, the position of peaks that potentially correspond to a wave need to be automatically extracted from the perturbation profiles. During flare/CME events, various localized changes in intensity and disturbances may occur, and it is not straightforward to identify a wave. Therefore, we make also use of the time information in subsequent images. First, the identification of peaks in perturbation profiles derived from a single frame are done by a peak finder. Thereafter, the peaks are checked across time, whether they follow a continuous outward motion (Section~\ref{subsec:wave_tracing}). As the results of the peak finder heavily influence the quality of the wave tracer, a custom peak finder was implemented, as it allows for a better control of the peak defining parameters.  

Before starting its main loop, the peak finder identifies all local peaks above a given threshold. This is done by the \textit{find\_peaks} function of the SciPy.signal library \citep{virtanen_scipy_2020}. The main loop of the custom peak finder starts with the largest peak that is found, and identifies its front and trailing edge given as percentage of the peak amplitude above the quiet level of $1.0$ (recall that we use base-ratio images). For the definition of the front and trailing edge, we use the full width at half maximum (FWHM), i.e. the crossing-points where the perturbation profile has decreased to $50$\% of its peak value. Alternatively, the FWHM can be also replaced by an absolute level. The front and trailing edge act as boundary of a peak, giving the range around the peak where no further peaks can be detected. In the next step, the algorithm searches the next highest peak outside these boundaries. For this peak, it checks for its prominence, i.e. the height difference between the peak and the deepest valley adjacent to the peak before an already verified peak, or the end/start of the profile. Both the prominence before and after the peak need to exceed a certain threshold for the peak to be considered. If the peak is considered, its front and trailing edge is derived (as described for the first peak) and the next lower peak outside the boundaries is investigated. The default minimum peak height we use is $1.03$ and the default prominence height is $0.03$. 

\begin{figure*}
    \centering
    \includegraphics[width=0.7\linewidth]{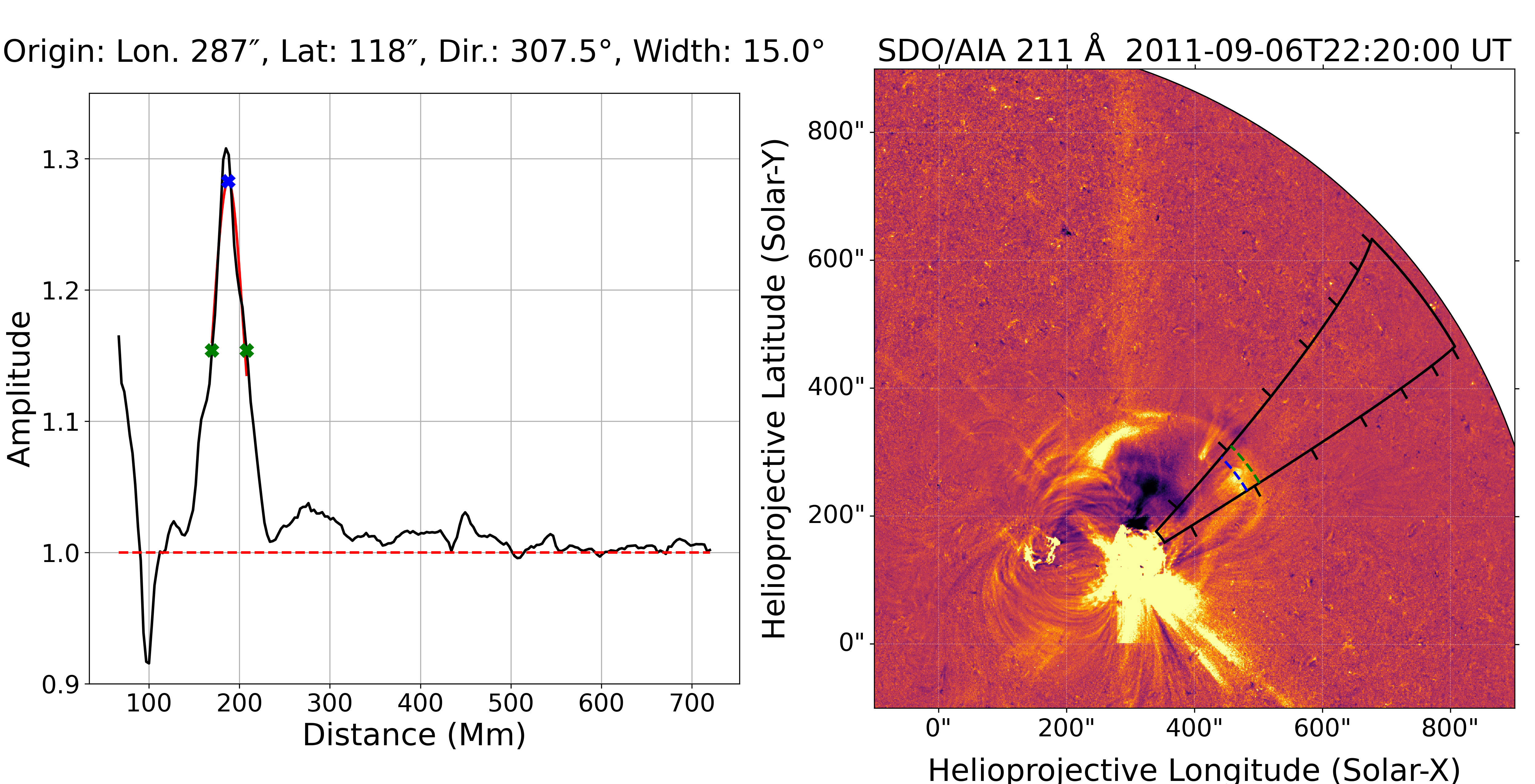}
    \includegraphics[width=0.7\linewidth]{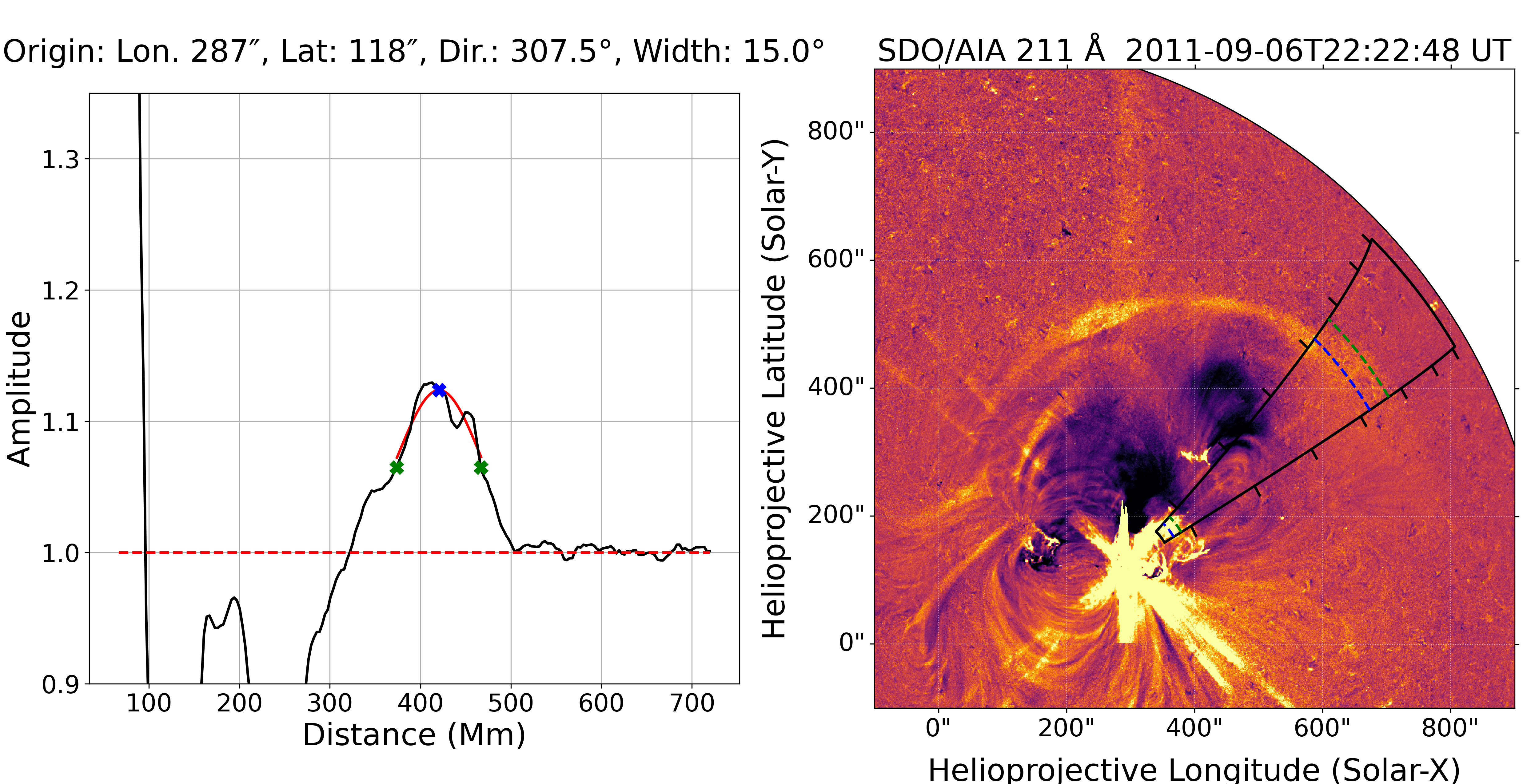}
    \includegraphics[width=0.7\linewidth]{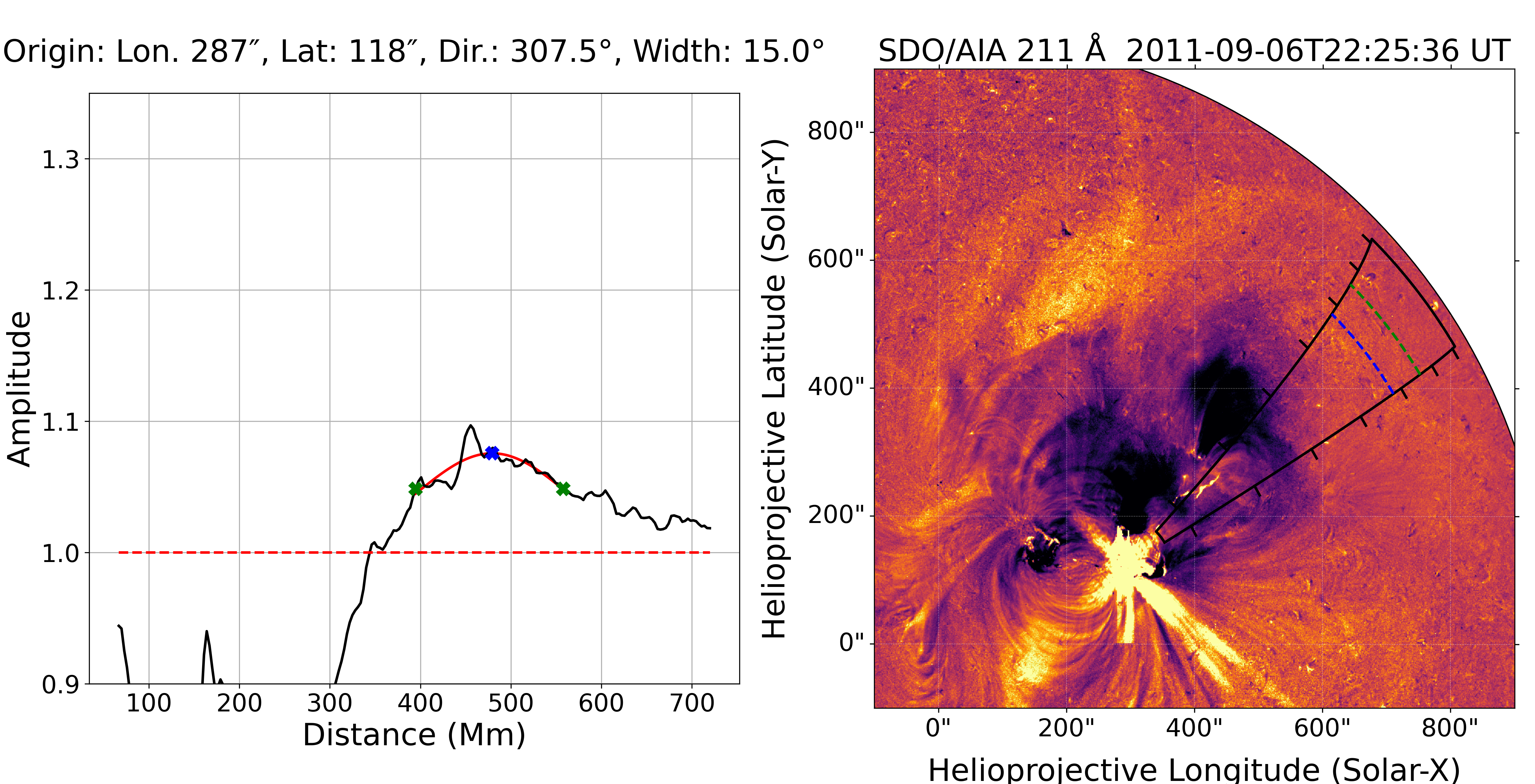}
    \caption{Demonstration of the Gauss fitting and peak finding of the perturbation profiles. The left panels shows the perturbation profile for three time steps. The red solid line shows the Gaussian fit to the measurements (black line), the red dashed line denotes the pre-event level. The  blue crosses indicates the peak identified by the peak finding algorithm, the green crosses the front and trailing edges (defined by the FWHM). The right panels shows the corresponding AIA $211$~\AA~base-ratio image in the range of [0.5,1.5]. Over-plotted is the investigated sector in the perturbation profiles, ranging from $-60$° to $-45$° (black lines). The markers on the sector show $100$~Mm increments. In the sector, the distance of the front and the peak of the wave as derived from the perturbation profiles plotted on the left are shown as green and blue dotted lines, respectively.  A movie of this plot is available in the online version over the time range 22:16 to 22:30 UT.}
    \label{fig:movie_frame_northwest_ward}
\end{figure*}

As the peak often does not follow the form of a Gaussian and since the profiles can be noisy, a robust estimate is desirable. This raises the need for a more robust peak position and amplitude detection. 
To this aim, we fit the peak only between the front and trailing edge with a Gaussian 
\begin{equation}
    f(x) = A \cdot \exp{\left(-\frac{(x-x_0)^2}{2\sigma^2}\right)},
\end{equation}
with $A$, $x_0$ and $\sigma$ as free parameters. As fitting algorithm the \textit{curve\_fit} function of \textit{scipy.optimize} \citep{virtanen_scipy_2020} is used with its default trf (Trust Region Reflective) algorithm using boundary conditions \citep{branch_subspace_1999}. The boundary conditions for $A$ and $\sigma$ are set from 0 to two times the maximum amplitude found within the fit regime and ten times the distance between the frontal and trailing edge respectively, large enough to not constrain the fit in these parameters. As boundary condition for the fitted peak position $x_0$, the front and trailing edge are used.  This allows for a good fit for both Gaussian shaped peaks as well as skewed ones, as the fitted Gaussian can move close to one boundary, if needed. Figure~\ref{fig:movie_frame_northwest_ward} shows a sequence of perturbation profiles over-plotted with the fitted peak, frontal and trailing edge along with the corresponding base ratio images, in which the distances derived from the perturbation profiles for the peak and the frontal edge are marked.


\subsubsection{Wave Tracing}
\label{subsec:wave_tracing}

The central part of the SOLERwave tool is the wave tracing algorithm, connecting the information of the peak finder to estimate wave speeds. It takes as input the distance versus time information of either the peak or the front determined by peak finder (the fitted peak distance or the front distance). It follows four main steps, iterating over all peaks (or fronts) in one timestep before continuing with the subsequent timestep. 

\begin{figure*}[h!]
    \centering
    \includegraphics[clip, trim=1cm 9cm 0cm 7cm,width=0.7\linewidth]{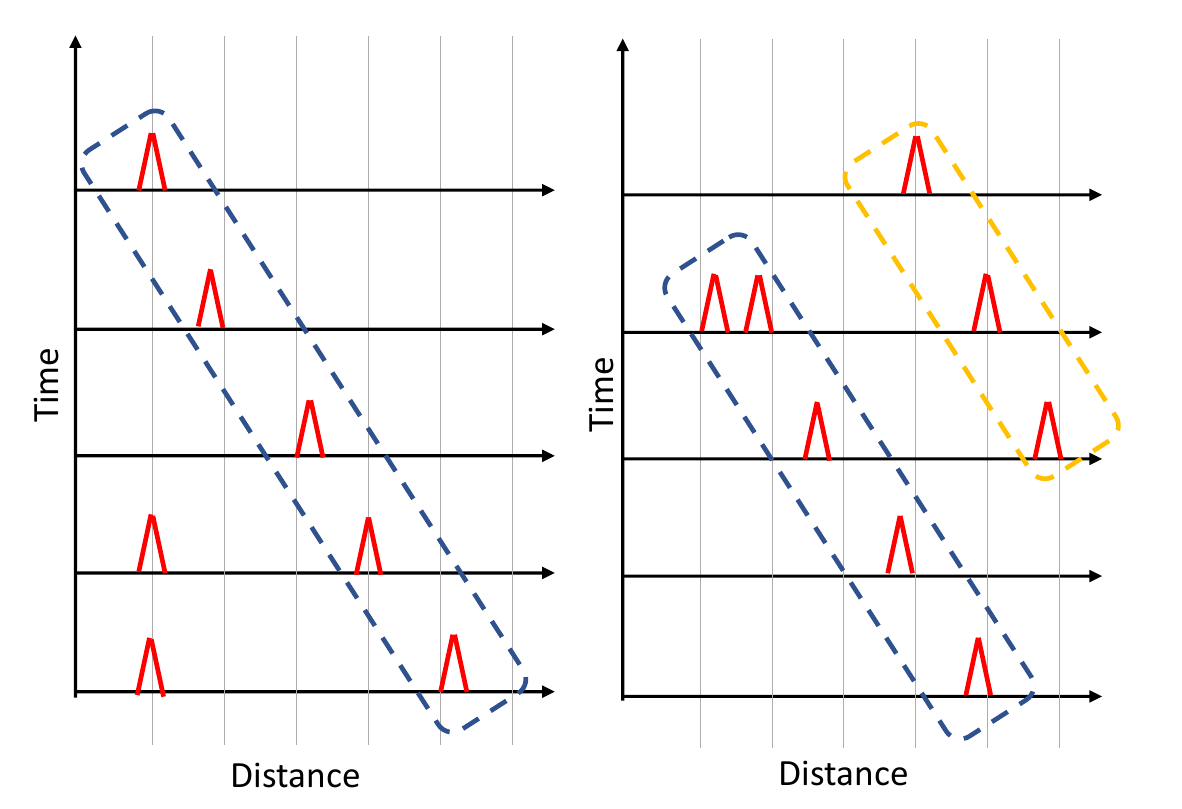}
    \caption{Schematic of the wave tracing algorithm. Left: The peaks circled in blue are found to be part of a wave by the tracing algorithms, as they follow each other within a minimum and maximum distance moving forward in time. The two peaks towards the bottom left did not move between time steps and are therefore not marked as a wave. Right: The blue dashed line circles an example for a two-peak start, where both peaks could be the correct starting peak. The orange dashed box circles a wave traveling further out in parallel. The algorithm can track multiple parallel waves. They do not interfere with each other as long as the peaks are separated by more than the maximum wave detection distance.}
    \label{fig:wave_tracing}
\end{figure*}

In step 1, the wave tracer compares a single peak at time $t_i$ with the peaks found at  $t_{i+1}$. All peaks fulfilling a certain minimum and maximum speed requirement (calculated as the distance between the peak at time $t_i$ and one peak at time $t_{i+1}$ divided by the corresponding time difference $t_{i+1} - t_i$), are marked. For this paper, a minimum speed of $100$~km~s$^{-1}$ and a maximum speed of $2500$~km~s$^{-1}$ is used, which lie well within the range of observed large-scale coronal wave speeds \citep{warmuth_kinematical_2011,nitta_large-scale_2013,muhr_statistical_2014}. In step 2, the algorithm checks if peaks marked as potential waves are already part of a wave. A peak-pair neither marked as part of a wave at $t_i$ nor $t_{i+1}$ is identified as a new wave by the algorithm. If the peak at $t_i$ is already part of a wave, the peak at $t_{i+1}$ is added to this wave. If the peak at $t_{i+1}$ is already part of a wave – marked as such by a different peak at time $t_i$ previously investigated – the algorithm checks whether the wave existed at $t_{i-1}$. If it did, the peak at $t_i$ is excluded. If not, it is added to the wave at $t_{i+1}$ as a dual start, a scenario where two peaks could technically be the start of the wave, so both are included. If both peaks (at $t_i$ and $t_{i+1}$) are already part of a wave, the wave association stays unaltered. Figure~\ref{fig:wave_tracing} shows an illustration for the different cases of the wave tracer.

In step 3, after evaluating all peaks, the wave tracing tool checks the duration of the waves found and omits all below a given threshold, typically requiring a wave peak to be detected in four consecutive time steps (equal to four consecutive image frames). In the last step, a linear fit is applied to the distance-time data of the peaks that are considered to be part of a wave in order to estimate the speed of the wave. 

Besides the primary wave feature, the wave fitting function allows also to apply a linear fit to a secondary feature, e.g., the wave front edge if the wave peak is traced, and vice versa. For the fit of a secondary feature the algorithm simply takes the distance values of the secondary feature and applies a linear fit to the distance-time data, without checking if they would fulfill any speed requirements themselves.

\subsection{Multi-sector Method and Huygens Plotting}

Besides investigating a wave in only one sector at a time, multiple sectors can be analyzed simultaneously  to gain information about the whole wave front. This is done by using the creation of perturbation profiles (Section \ref{sec:Creating Perturbation Profiles}) and identifying and fitting of peaks (Section \ref{sec:Peak finding and fitting}) of the single-sector algorithm on multiple sectors of $10$° with an overlap of $9$° (i.e. the mean direction of the sectors is separated by $0.5$°). To ensure the detection of smooth wave fronts independent of the different peak heights in different directions we use for the identification of the wave front a fixed absolute value of $1.02$. This produces a map full of fronts tracing the wave at time $t_i$. To infer motion, the distance between each front-distance at $t_i$ against all front-distances from the previous timestep $t_{i-1}$ is calculated. At each time step, the distance calculation follows great arcs, i.e the shortest distance between two points on a sphere. Then the local speed is derived from the distance between these two points and the time between the two observations. As with the wave tracing in the single-sector analysis, this speed is checked against a minimum value of $100$~km~s$^{-1}$ and a maximum value of $2500$~km~s$^{-1}$. If the value is within the range, the velocity vector pointing from the point at $t_{i-1}$ to the point at $t_i$ is stored and later displayed in a quiver plot. Further, the angle of the 2D velocity vector from the radial great arc with respect to the presumed wave origin is calculated. As the distance uncertainty increases steeply for waves having been refracted significantly from the radial direction, local wave arrows exceeding $\pm 45$° from the radial direction are excluded.

\subsection{Uncertainty Analysis}

SOLERwave incorporates an uncertainty analysis and propagation. In this subsection, we give a concise synopsis of the components. A comprehensive derivation of the various terms can be found in the Appendix. The different error terms include the amplitude uncertainty within a segment, which is translated to a distance uncertainty. To this the errors arising from the finite size of pixels and segments are added. The amplitude uncertainties are used as relative weights for the Gaussian peak fit and and distance uncertainties as relative weights for the linear fit of the time-distance data in order to derive the wave speed. The uncertainties of the fit parameters are estimated from their covariance matrices. Note that the uncertainties given are often an upper limit. 


Uncertainties not taken into account in the kinematic plots are the uncertainty of the wave origin  (resulting in an increased speed if the wave does not propagate perfectly circular outward), and the error induced by the correction for solar differential rotation (which is very small over the considered time frames of some ten minutes). Further not included is the large, but highly correlated distance uncertainty arising from the propagation height of the wave. The challenge in precisely quantifying the propagation height of large-scale coronal waves stem from the fact that the observed intensity is a line-of-sight integration of the optically thin emission over varying propagation heights, and depends also on the observer viewpoint \citep{hoilijoki_interpreting_2013,downs_validation_2021}. Previous studies that used tomographic reconstruction and limb events \citep{kienreich_stereo_2009, patsourakos_what_2009,delannee_time_2014, podladchikova_three-dimensional_2019,hou_three-dimensional_2022} estimate the travel height of the upper bound of the visible front of large-scale coronal waves at $10$~Mm to $100$~Mm. The projection effects of waves traveling at these heights onto the solar surface lead to significant correlated distance uncertainty (see further details in Appendix~\ref{appendix:sec_Projection_Uncertainty}).

\subsection{Implementation and Tool Availability}

The SOLERwave tool and its accompanying file loader were developed in Python 3.10. For manipulation and loading of the data, these tools use the SunPy library \citep{sunpy-community_sunpy_2023}, which itself is heavily based on the Astropy library \citep{astropy_collaboration_astropy_2022}. The tool is based on Jupyter Notebooks. The accompanying file loader allows for download and preprocessing of data using a single function, internally using the SunPy \textit{Fido} function. The publicized versions of the tool including the single-sector analysis and the file loader is available on Github \url{https://github.com/soler-he/}. The standard graphic output of the single-sector analysis include the overview plot with J-maps in $8$ sectors of $45$° width (``octant plot"), the time evolution of the perturbation profiles in a sector selected, and the final kinematics plots. All of these output plots will be applied and shown in Section~\ref{sec:results}. Further a movie output is created of the evolution of the images and perturbation profiles. 

The SOLERwave tool was created as part of the Europe Horizon project SOLER, which aims to investigate energetic solar eruptions from three perspectives: fast coronal mass ejections (CMEs), strong X-ray flares, and large solar energetic particle (SEP) events \citep{vainio_horizon_2025}. Because of their link to particle energisation (e.g. \citealt{dresing_17_2023}), the project studies also large-amplitude coronal waves and shocks.

\section{Results}
\label{sec:results}

In the following we use the SOLERwave tool to analyze the fast and complex large-scale coronal wave observed at 22:20~UT on September 6, 2011. This wave has been already investigated in previous studies, where a mean speed of $1246$~km~s$^{-1}$ \citep{nitta_large-scale_2013} and $900\pm30$~km~s$^{-1}$ to $1070\pm70$~km~s$^{-1}$ \citep{dissauer_projection_2016} was derived. Also, as was found in \citet{dissauer_projection_2016}, the wave showed a transient disappearance and reappearance in the AIA $211$~\AA~filter, which was interpreted as an effect of different structures along the line of sight. In this paper, we concentrate in particular on the varying wave speeds in the different propagation directions. As in \citet{nitta_large-scale_2013} and \citet{dissauer_projection_2016}, the flare position at $14$° North, $18$° West will be used as presumed wave origin.

\begin{figure*}
    \centering
    \includegraphics[clip, trim=1.75cm 2.75cm 2.25cm 1cm,width=1\linewidth]{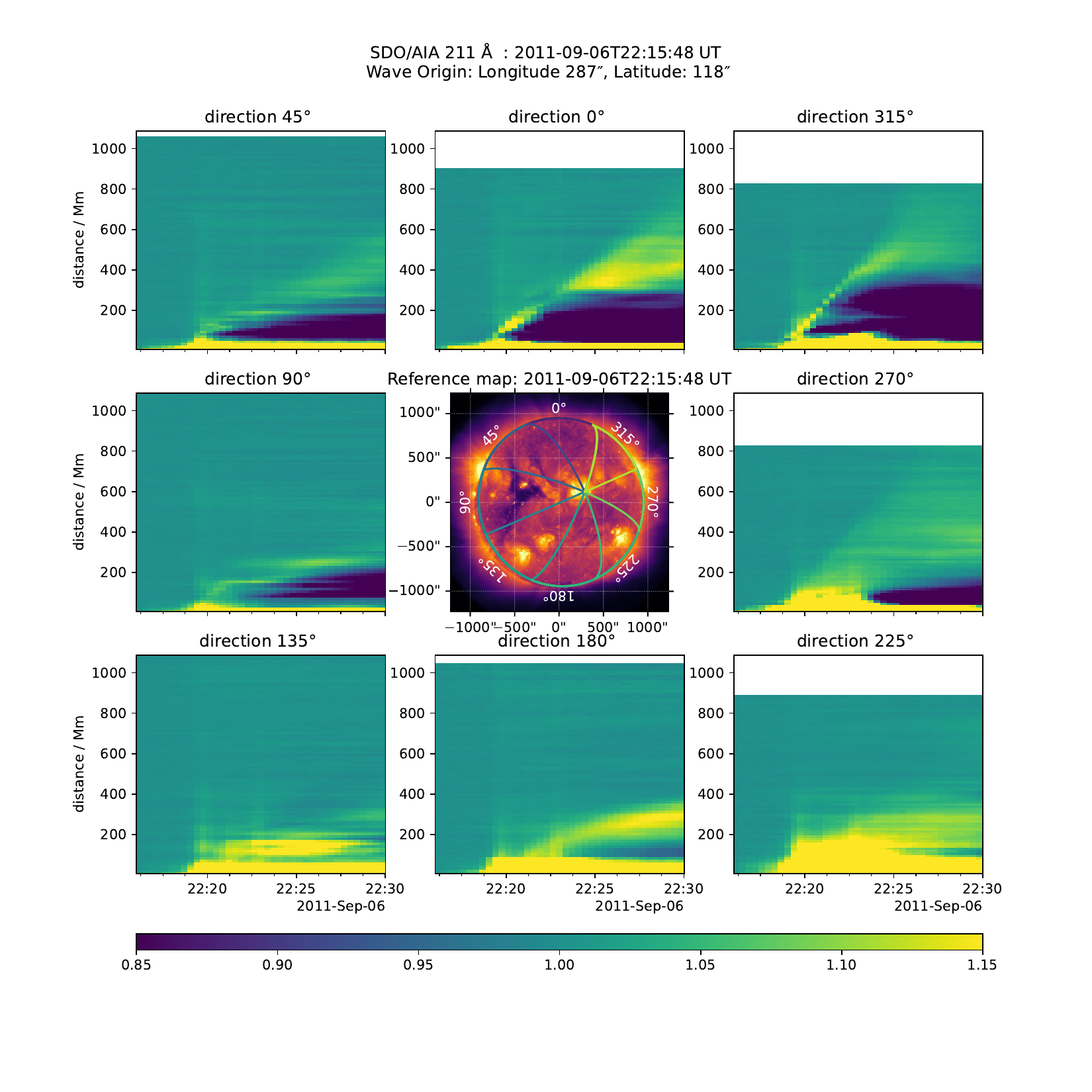}
    \caption{Octant plot for the large-scale coronal wave on September 6, 2011 showing the wave evolution in SDO/AIA $211$~\AA~in eight sectors, each with a width of $45$° (the sector boundaries are illustrated in the base image shown in the middle panel). The directions are counted from solar North ($0$°) in counter-clockwise direction. The color bar below quantifies the corresponding intensity values from base ratio images (which are dimensionless), which we restricted in the visualization to the range [$0.85$,$1.15$]. The white areas visible in the directions $315$° and $270$° for large y-values are the result of these areas being not visible to the observer. The very bright, yellow regions at low y-values are due to the associated flare.}
    \label{fig:octantplot}
\end{figure*}

Figure~\ref{fig:octantplot} shows the octant plot, the first step in the analysis. It is used to obtain an overview on the wave evolution/strengths in the different directions, and to choose the sectors in which the wave will be studied in the single-sector analysis. The octant plot consists of eight so-called J-maps derived from perturbation profiles in the different directions. J-maps are plots with time on the $x$-axis, distance on the $y$-axis and the color encoding intensity (from base ratio images). In these plots, the propagating wave appears as an oblique line of enhanced emission. Each J-map has been created from perturbation profiles calculated along sectors with a width of $45$° and a longitudinal resolution of $1$°. It is similar in appearance as the plot used by \citet{nitta_large-scale_2013}. The octant plot shows the propagation of the wave front, which is strongest towards North and Northwest, in agreement with previous findings \citep{nitta_large-scale_2013,dissauer_projection_2016}.

\subsection{Single-Sector Analysis}
\label{sec:sing_sector}

   \begin{figure*}[!h]
        \centering
        \includegraphics[width=1\linewidth]{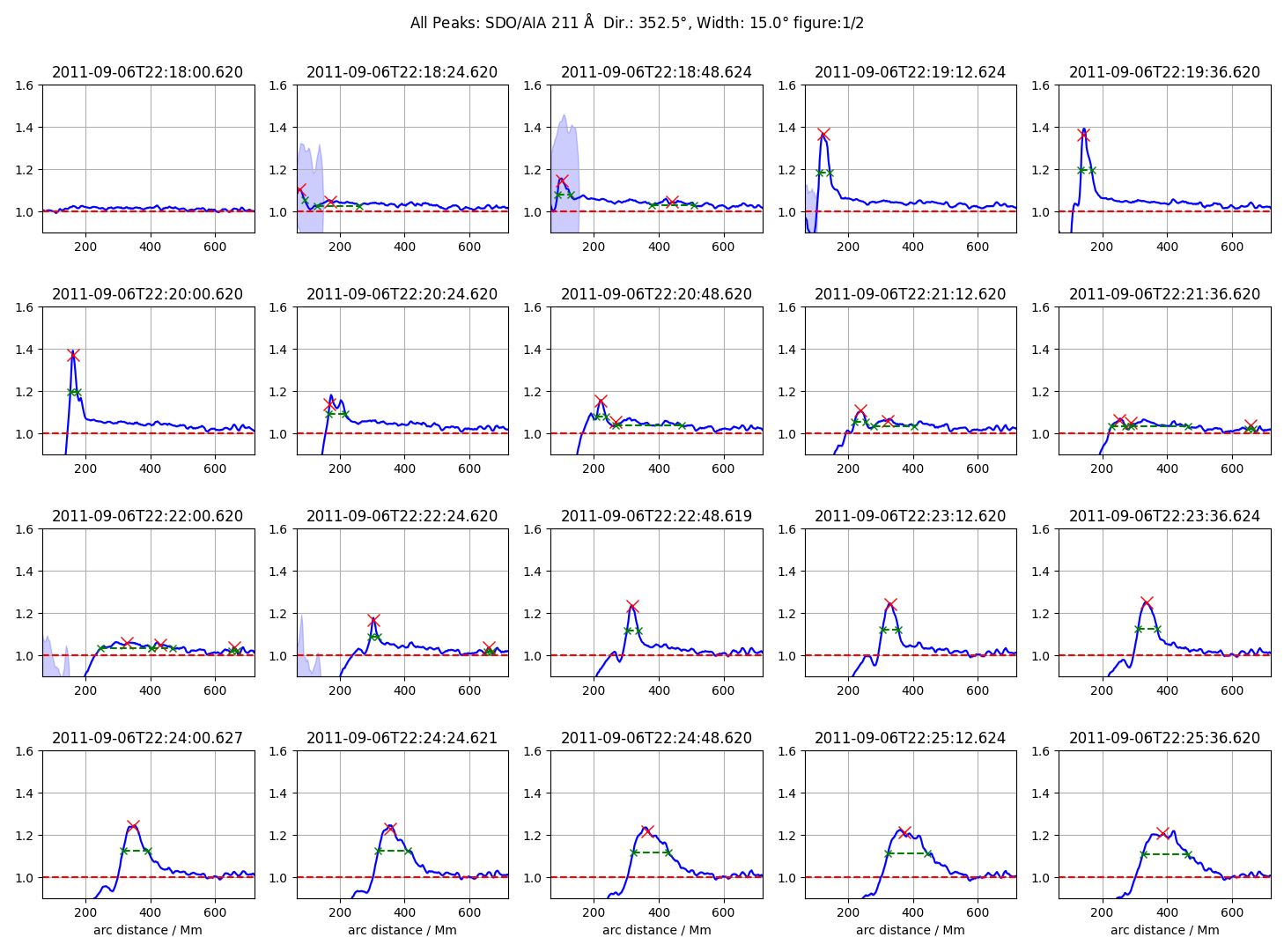}
        \caption[Perturbation profiles for the northward wave]{Perturbation profiles derived from SDO/AIA $211$~\AA~base ratio images for the wave observed on September 6, 2011 along the sector between $-15$° to $0$° (North; see also Figure~\ref{fig:movie_north_wave}) for different time steps.  The blue line is the perturbation profile derived, with the blue shaded area the segment standard deviation. The red crosses mark the position of the peak of the fitted Gaussian function (not shown in this plot) found by the peak finding algorithm, the green crosses the front and trailing edge of each peak calculated at FWHM. The red dashed line denotes the pre-event level. Shown are only the first 19 time steps after the first wave detection, not the full time range. }
        \label{fig:perturbation_profiles}%
    \end{figure*}

We now focus on two primary directions to use the single-sector analysis, namely the propagation of the wave toward the North and toward the Northwest. For the analysis of the northward traveling wave, a sector of $15$° width toward the solar North is chosen (sector boundaries are $-15$° and $0$° with respect to the great circle intersecting both the wave origin and the North, shown in Figure~\ref{fig:movie_north_wave} and the associated movie). Figure~\ref{fig:perturbation_profiles} shows the perturbation profiles observed along this sector between 22:18:00~UT and 22:35:36~UT. The first three panels show blooming effects due to the strong flare emission into the sector of observation at small distances, leading to large uncertainty bands (blue shaded area). At 22:19:12~UT, the first bright wave front can be seen propagating towards the North, observed as sharp peak in the perturbation profiles reaching a maximum amplitude of $1.37$ at 22:20:00~UT. This bright front decreases in amplitude in the next minutes until 22:22:00~UT, before rising again to over $1.2$ within one minute. The last five panels (from 22:24:00~UT) show the wave pulse broadening as it gets refracted toward the North, an effect also seen in Figure~\ref{fig:movie_north_wave} and the accompanying movie.

\begin{figure}
    \centering
    \includegraphics[clip, trim=1.25cm 1.5cm 1.75cm 1cm,width=1\linewidth]{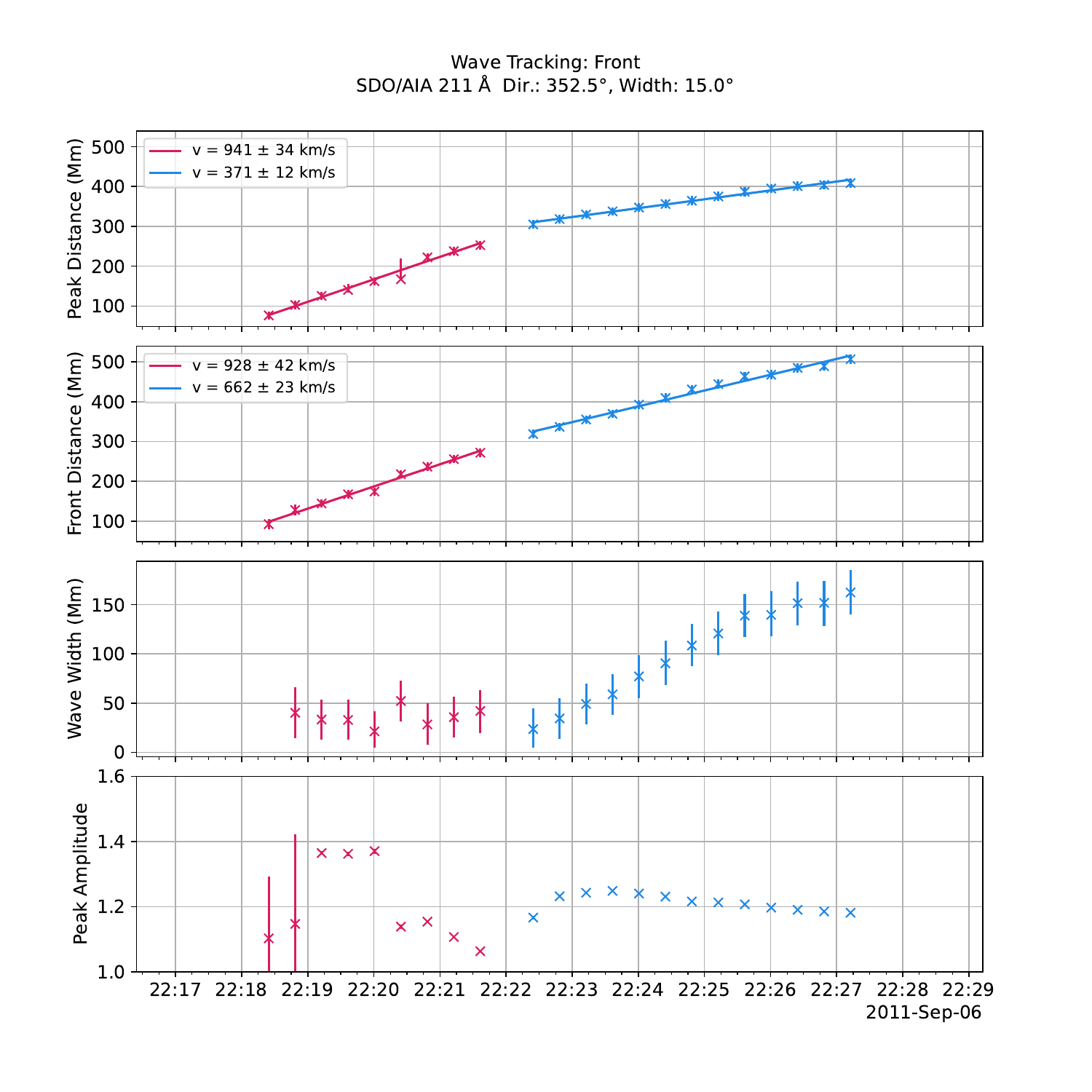}
    \caption{Kinematics of the wave in the north direction, as provided by the SOLERwave output plots. From top to bottom: distance of the peak of the wave, the front of the wave, the wave width, and amplitude. Shown are only points associated to waves detected by the wave tracing algorithm. The color shows the association to a continuous wave detection. Two waves have been detected, which can be assumed being part of one continuous wave traveling north. }
    \label{fig:kinematics_northward}
\end{figure}

Figure~\ref{fig:kinematics_northward} shows the kinematics plot derived from the perturbation profiles. In order to increase the reliability of the wave tracing algorithm in this particular study, the front edge of the wave is tracked instead of the peak (as was described as an option of the algorithm in Section~\ref{subsec:wave_tracing}). The upper two panels show the distance of the peak and the front of the wave, derived from the Gaussian fits to the profiles. The wave tracer finds two waves, one (shown in red) for the initial phase of the wave with a linear speed of $940 \pm40$~km~s$^{-1}$ for the peak and  $930 \pm50$~km~s$^{-1}$ for the front of the wave. The second wave (in blue) covers the later parts of the wave with a linear speed of $370 \pm 20$~km~s$^{-1}$ for the peak and  $660 \pm 30$~km~s$^{-1}$ for the wave front. The separation into two instead of one continuous wave detection is due to the transient vanishing of the wave front below the noise level at 22:22:00~UT (see the gaps in Figure~\ref{fig:kinematics_northward}).

The bottom two panels of Figure~\ref{fig:kinematics_northward} show the wave width and peak amplitude of the Gaussian fits. While for the initial phase (plotted in red) a relative constant wave width can be observed, the later phase (plotted in blue) shows a nearly linear increase in wave width, which is a typical sign of wave broadening during the propagation. Note that the plot of the wave width misses the first data point, due to the wave intersecting the lower bound of the sector investigated above the FWHM of the respective peak, resulting in no trailing edge found.  The wave amplitude of the initial phase first steepens from $1.2$ to almost $1.4$ before decaying to below $1.2$ at around 22:21~UT. For the later phase the amplitude decrease slowly from above $1.2$ to below $1.2$ after reemerging from the noise floor at 22:22:24~UT, i.e. stays at relatively high amplitudes. 

\begin{figure}
    \centering
    \includegraphics[clip, trim=1.25cm 1.5cm 1.75cm 1cm,width=1\linewidth]{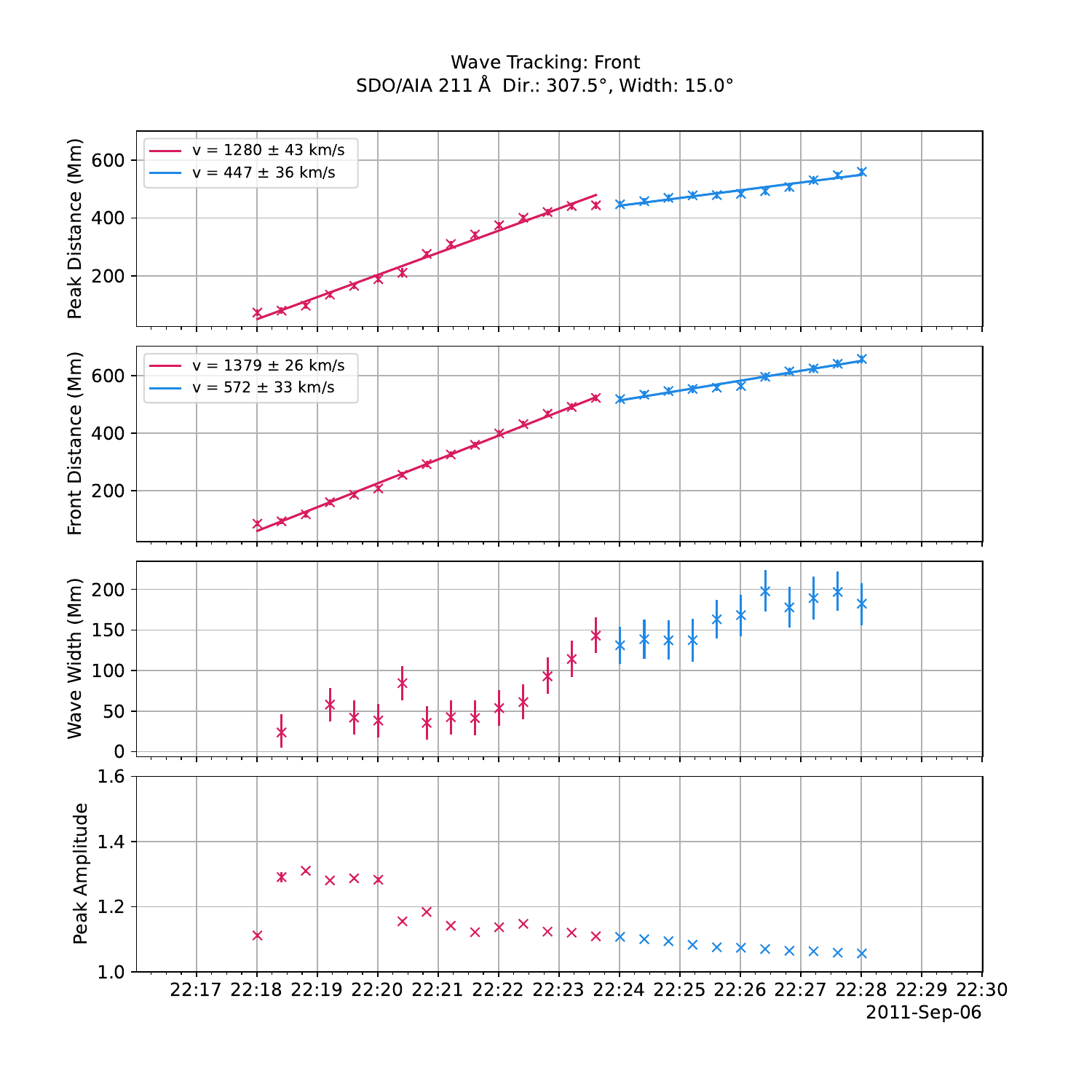}
    \caption{Same as Figure~\ref{fig:kinematics_northward}, but for the wave propagation in the northwest direction. Again, the algorithm splits the wave into two single waves (identified by red and blue colors)}
    \label{fig:kinematics_northwest_ward}
\end{figure}

For studying the wave in the northwest direction, a $15$° sector with boundaries at $-60$° and $-45$° was chosen (shown in Figure~\ref{fig:movie_frame_northwest_ward}). Figure~\ref{fig:kinematics_northwest_ward} shows the resulting kinematics plot. In this case, the SOLERwave tool identified again two waves: A faster one at the start and a slower one towards the end. For the initial phase, the speed derived from the linear fit is $1280 \pm50$~km~s$^{-1}$ for the peak of the wave and $1380 \pm30$~km~s$^{-1}$ for the front of the wave. In the later phase the fit results in $450 \pm40$~km~s$^{-1}$ for the wave peak and $570 \pm40$~km~s$^{-1}$ for the wave front. Checking the wave front for continuation using the movie of Figure~\ref{fig:movie_frame_northwest_ward}, we find indeed only one bright front moving outwards. The separation into two instead of one continuous wave detection is due to the stagnation of the wave front distance around 22:24:00~UT, not fulfilling the minimum distance requirements for two consecutive points of a wave. The trend of the wave width in the kinematics plot in the northwest direction shows an increase in wave width for the full wave observation, while the trend of the  peak amplitude shows an initial increase (wave steepening) to an amplitude of $1.3$, followed by a continuous decrease after 22:20~UT. 

We note that large-scale coronal waves often show a speed that is higher than the fast magnetosonic speed of the corona and that decreases during its evolution, in association with a decrease in wave amplitude (e.g. \citealt{warmuth_kinematical_2011,muhr_statistical_2014}). This is due to the nonlinear nature of large-amplitude or shock waves, in which the propagation speed is dependent on the amplitude of the wave \citep{mann_simple_1995,vrsnak_origin_2008}. In our analysis, the automatic algorithm erroneously splits the wave into two single waves. The distinct difference in the two linear speeds derived (the latter segment always being substantially slower) can be taken as a reflection of the decrease of the wave speed during its evolution, in accordance with the decreasing wave amplitude. In addition, the wave and the CME front can often not be distinguished especially at the beginning of the evolution (the wave has not yet separated itself from the driver) and by using observations from only one single viewpoint \citep{dissauer_projection_2016}.

\subsection{Multi-Sector Analysis}

   \begin{figure*}[!h]
        \centering
        \includegraphics[width=1\linewidth]{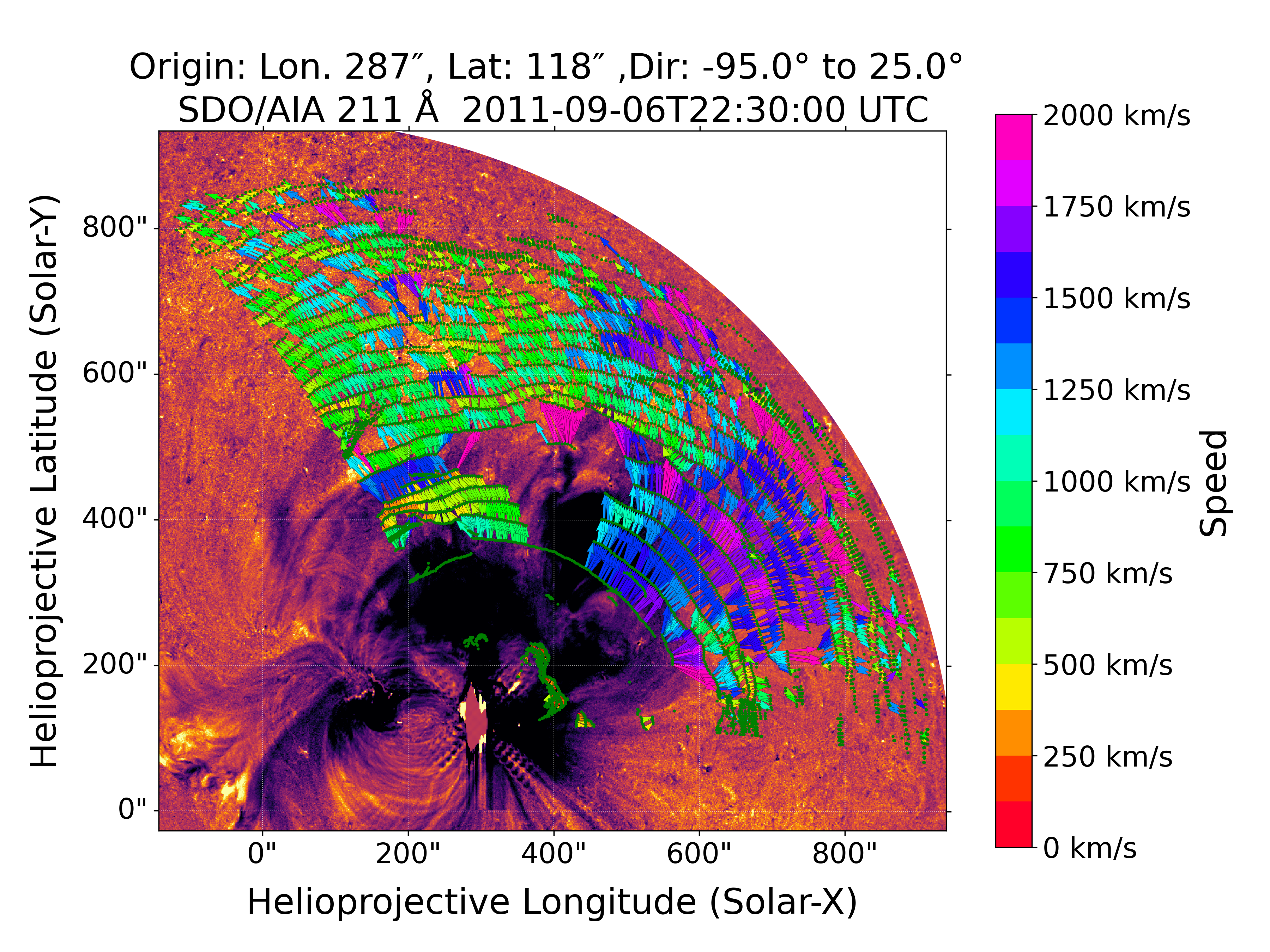}
        \caption{Multi-sector analysis in the SDO/AIA 211\AA~passband: velocity of  wave front. The arrows connect the closest wave front between time steps, the color shows the velocity based on the distance and time difference between observations. The shown map is the result of all time steps. A movie of this plot is available in the online version showing the stepwise creation.}
        \label{fig:Huygensplot_Velocities}%
    \end{figure*}


\begin{figure}[!h]
    \centering
    \includegraphics[width=1\linewidth]{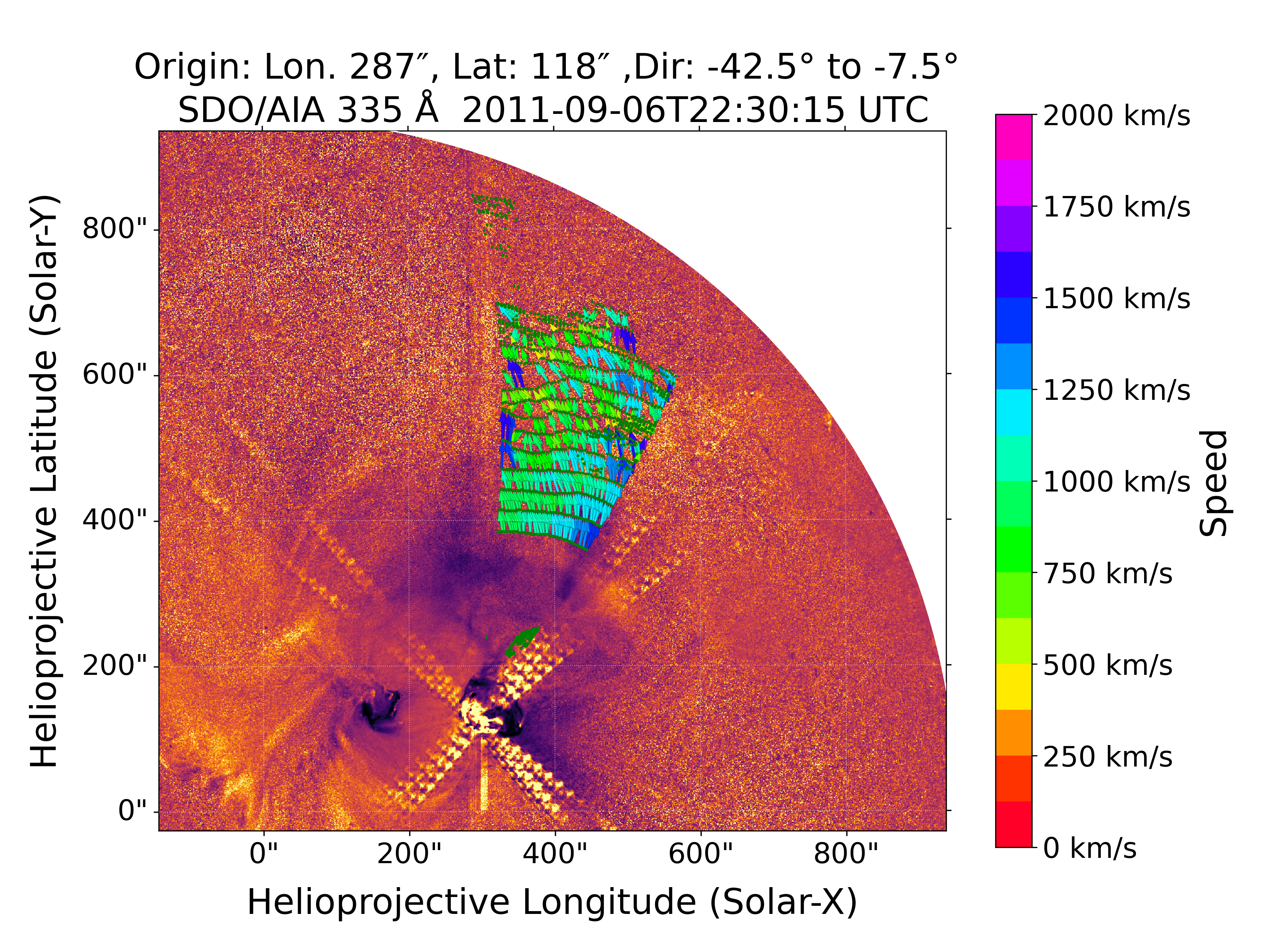}
    \caption{Same as Figure~\ref{fig:Huygensplot_Velocities} but for the SDO/AIA $335$~\AA~passband.}
    \label{fig:Huygensplot_Velocities_335}
\end{figure}

To investigate the speed difference found in the single-sector analysis and to investigate the 2D velocity vector, the multi-sector analysis was performed. Figure~\ref{fig:Huygensplot_Velocities} shows the result in a sector between $-95$° and $25$° with respect to the wave origin and the solar North. The plot shows multiple notable features of the wave. First, a distinct dichotomy between the speed in the north and west direction can be observed. While the velocities towards the North are between $750$ km~s$^{-1}$ and $1000$ km~s$^{-1}$ the velocities towards the West are significantly higher ranging from $1250$ km~s$^{-1}$ to $1750$ km~s$^{-1}$. Further, the wave is refracted towards the North. In general, it appears as if the wave front as a whole is tilted in a counter-clockwise direction. This indicates that the center of the wave front after 22:19:38~UT (the start time of the multi-sector analysis) is shifted from the flare position. Rerunning the multi-sector analysis with a limited sample of other wave origins shows that indeed, a wave origin at $14$°N, $24$°W (equivalent to a Helioprojective Longitude of $378^{\prime\prime}$ and Latitude of $12^{\prime\prime}$) reduces the amount of deviation from the radial direction already significantly. Lastly, the multi-sector plot shows a large patch without traced velocities at around Helioprojective Longitude $450^{\prime\prime}$ and Latitude $450^{\prime\prime}$, as the wave amplitude in this area dips below the minimum detection threshold for the wave amplitude of the algorithm. This was already reported by \citet{dissauer_projection_2016} as a specificity of this large-scale coronal wave event. 

To check if the difference in velocities transitions smoothly in this region we performed a multi-sector analysis of  SDO/AIA $335$~\AA~images, where the wave does not show the intermediate disappearance as in the SDO/AIA $211$~\AA~passband. Due to the lower signal and the resulting higher noise level on this filter, the minimum peak height was increased to $1.15$ and the amplitude at which we determine the front height to $1.1$. Figure~\ref{fig:Huygensplot_Velocities_335} shows the area in question and indeed a smooth transition between the lower velocity of  $750$ km~s$^{-1}$ towards the North an the higher velocity of $1250$ km~s$^{-1}$ towards the West can be seen. The velocities shown are limited to the area between $-42.5$° and $-7.5$°, as diffraction patterns of the flare and blooming of the CCD away from the flare region in the image interfere with the multi-sector analysis.

\section{Discussion and Conclusion}
\label{sec:discussion}

The fast and complex large-scale coronal wave observed on September 6, 2011 provides us with a unique opportunity to use the capabilities of the newly developed SOLERwave tool to investigate the velocity differences of the wave front and its evolution. For the initial phase of the wave we observe a speed of the wave front of $930\pm50$~km~s$^{-1}$ towards the northern direction and an substantially higher speed of  $1380\pm20$~km~s$^{-1}$ in the northwest direction. The derived speeds agree with previous studies of the event.  \citet{dissauer_projection_2016} report a speed of $900\pm30$~km~s$^{-1}$ for the peak towards the solar North. \citet{nitta_large-scale_2013} gives a maximum speed of $1246$~km~s$^{-1}$ for the wave, but do not mention the direction of measurement. As we find the maximum speed in the northwest direction, we presume \citet{nitta_large-scale_2013} to have found the given speed in this direction as well. In principle, speed differences like this could be explained by differences in the magnetosonic speed of the medium caused by e.g. active regions or coronal holes \citep{gopalswamy_euv_2009,liu_impacts_2018}, differences in the speed and/or effectiveness of the driver accelerating the wave \citep{kienreich_stereo_2009, veronig_genesis_2018} or by projection effects leading to systematic errors in the derived wave speed of the wave speed \citep{hoilijoki_interpreting_2013,podladchikova_three-dimensional_2019,downs_validation_2021}. 

\begin{figure}
    \centering
    \includegraphics[clip, trim=4cm 0cm 7cm 2cm,height = 4.9cm]{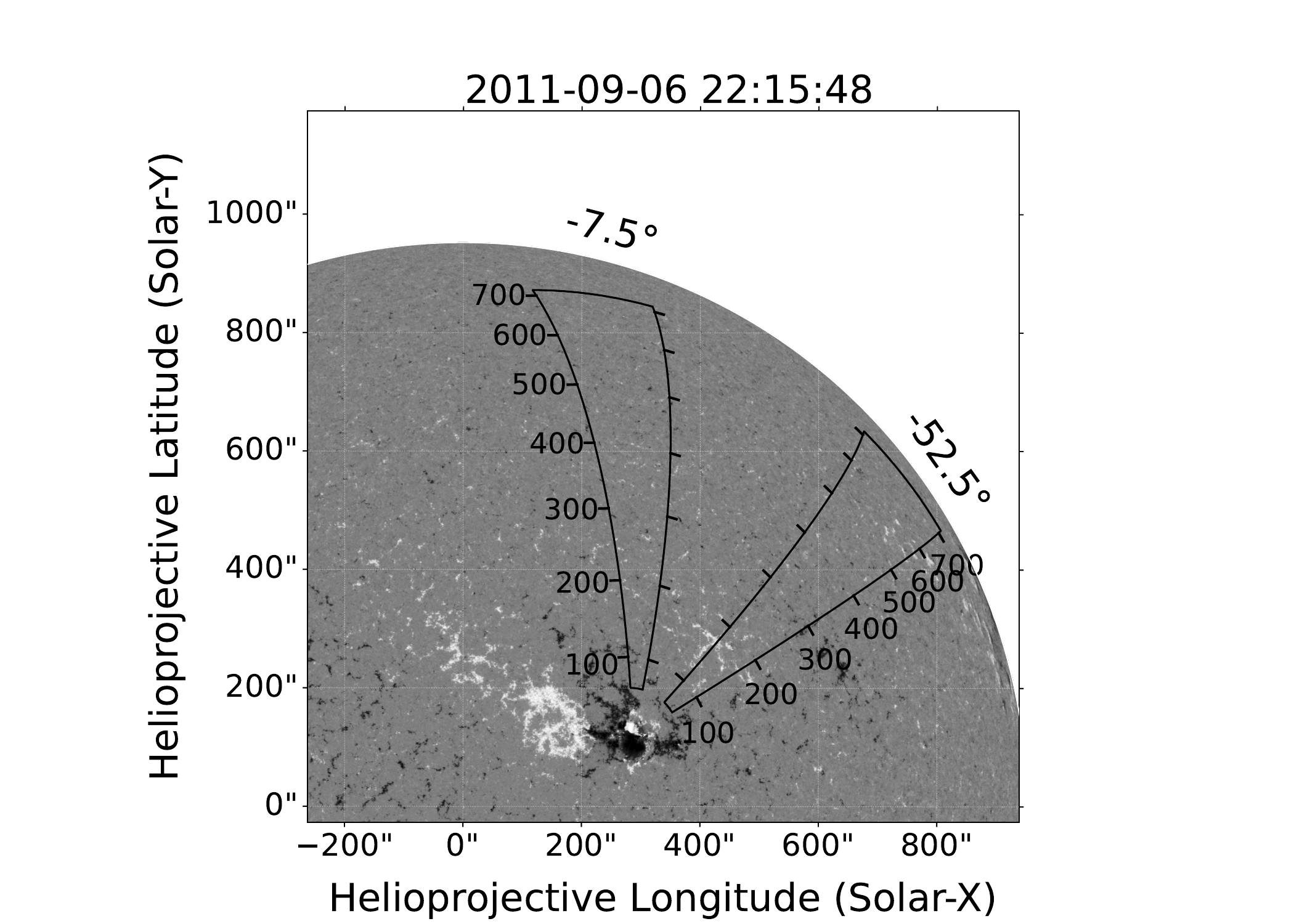}
    \includegraphics[clip, trim=5.5cm 0cm 0cm 2cm,height = 4.9cm]{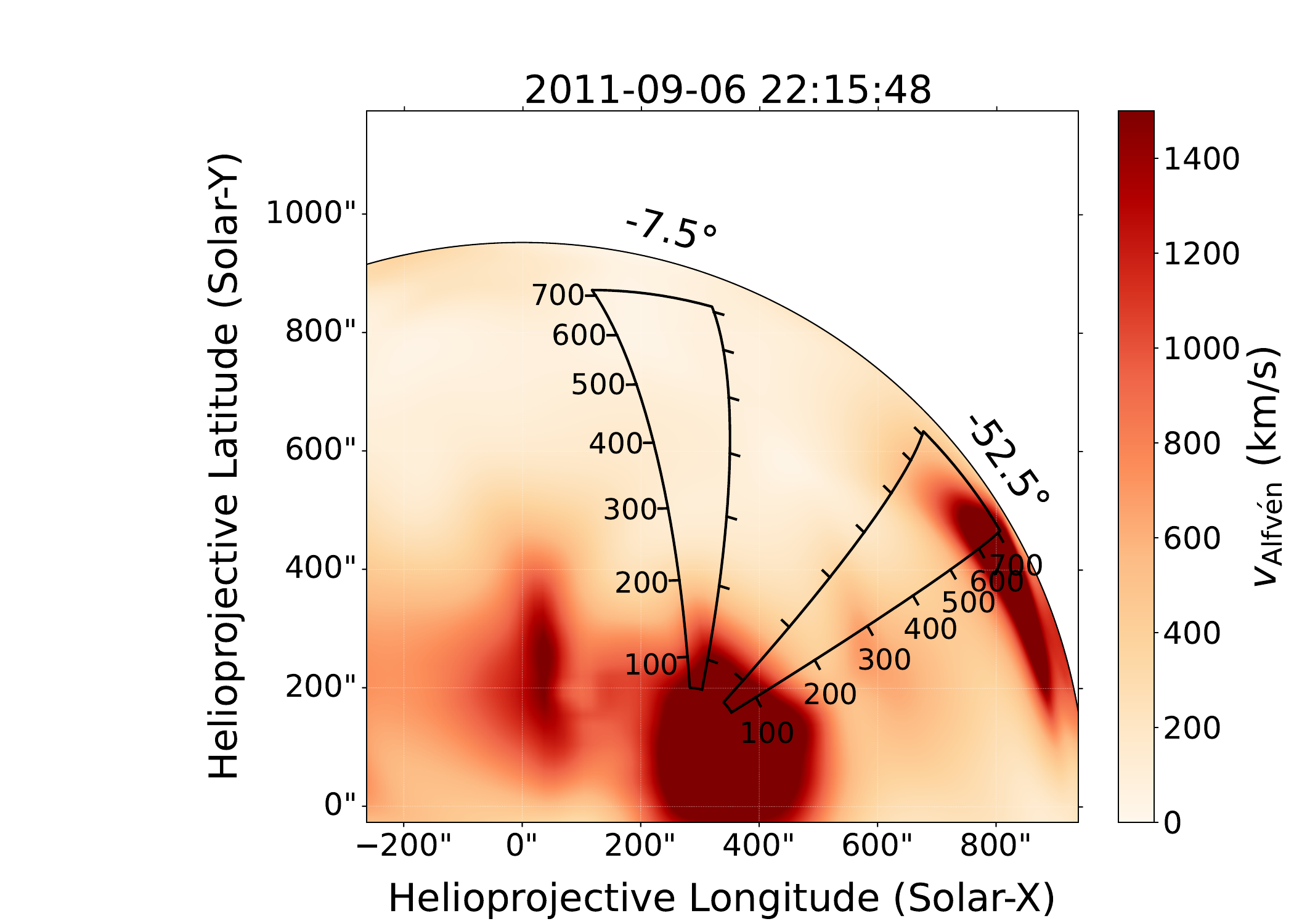}
    \caption{The left plot shows the SDO/HMI LOS magnetogram observed at 22:00:00~UT in the range of [$-1000$, $1000$] Gauss. The right plot shows a map of the Alfv\'{e}n speed at a height of $1.07$~$R_\odot$, derived from a thermodynamic MAST model for Carrington rotation 2114.  Overplotted are the sectors towards the north and northwest as used in this study. }
    \label{fig:mag_alphene_map}
\end{figure}

In order to investigate the spatial distribution of the magnetic field, as one key factor determining the local Alfv\'{e}n speed and magnetosonic speed, we extract the mean unsigned line of sight (LOS) magnetic field from magnetograms observed by the Helioseismic and Magnetic Imager(HMI) onboard SDO at 22:00:00~UT by applying the masks used to create the perturbation profiles. We ignore pixels with an absolute field strength below the noise level of $10$~G and filter the resulting profile with running average of $10$° window-size. In addition, to estimate the Alfvén speed in the ambient corona, we use a steady-state coronal magnetohydrodynamic (MHD) solution produced by the MAS (Magnetohydrodynamic Algorithm outside a Sphere) model \citep{lionello_multispectral_2009} developed by Predictive Science Inc. For the results presented here, we use the thermodynamic MAS (MAST) model employing a processed synoptic HMI magnetogram for Carrington rotation 2114 (specifically, we use the cr2114-medium/hmi\_mast\_mas\_std\_0101 run available from \url{https://www.predsci.com}). The model provides the MHD variables on a grid, from which we compute the Alfv\'{e}n speed as $v_A = B/\sqrt{\mu_0 \rho}$ using the SHELVIS visualization tool\footnote{https://github.com/jpomoell/shelvis}. For the evaluation we take a snapshot of the Alfv\'{e}n speed at a height of $1.07$ solar radii ($\approx 50$~Mm above photospheric height) and derive the mean value along the sectors using the masks used to create the perturbation profiles. Figure~\ref{fig:mag_alphene_map} shows the signed photospheric LOS magnetic field and the Alfv\'{e}n speed at a height of $1.07~R_\odot$, overlaid with the sectors in the north and northwest direction (with the sectors centered at $-7.5$° and $-52.5$° from solar North, respectively). Notably, while we can observe increased magnetic flux toward the northwest direction coinciding with an elevated Alfvén speed, the north direction appears magnetically quiet.

\begin{figure}
    \centering
    \includegraphics[clip, trim=0.5cm 0cm 0cm 1cm,width=0.9\linewidth]{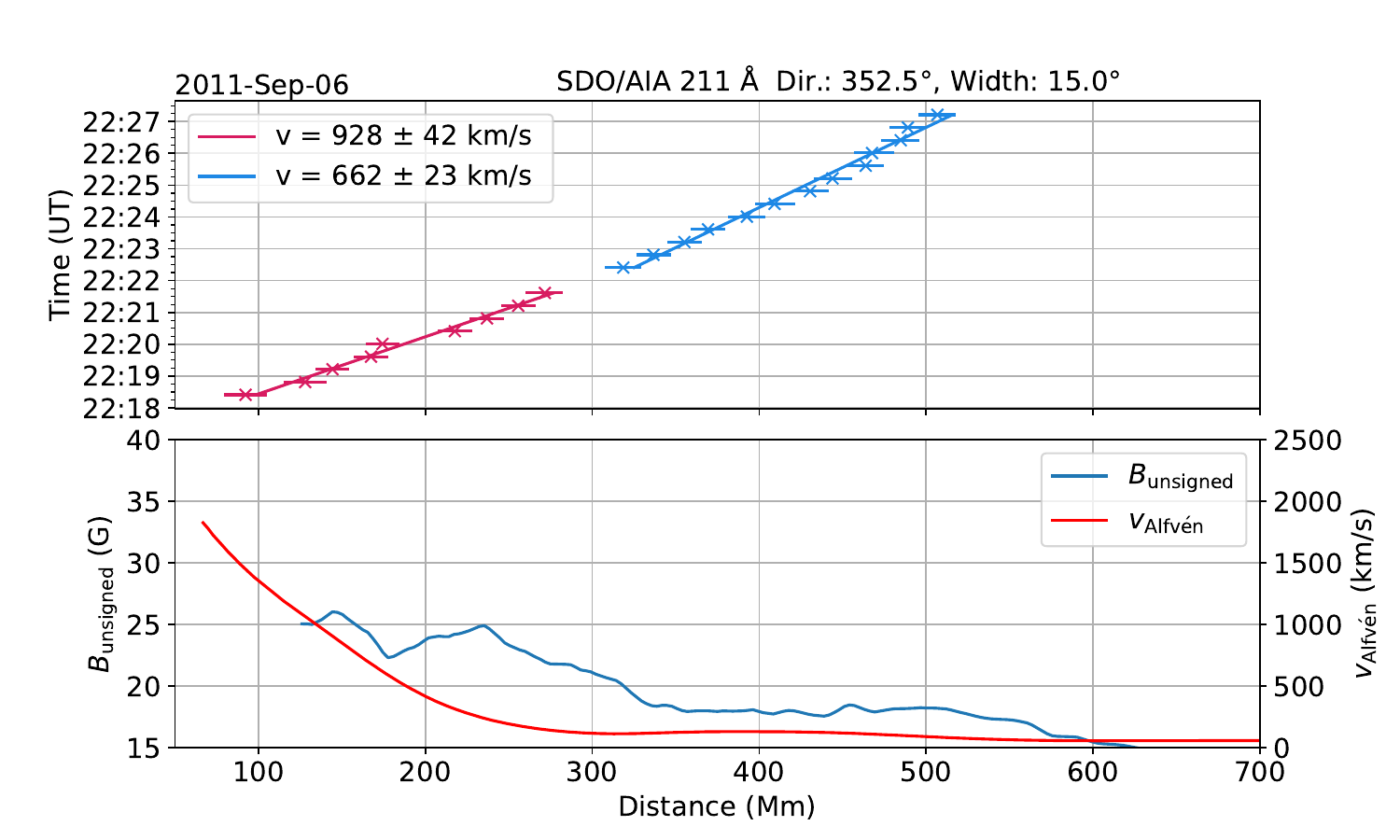}
    \caption{Kinematics of the wave front in the north direction in the upper panel (same as Figure~\ref{fig:kinematics_northward}) compared to the unsigned SDO/HMI LOS magnetic field and Alfv\'{e}n velocity simulated at a height of $1.07~R_\odot$ ($\approx 50$~Mm above the photosphere) in the lower panel.}
    \label{fig:mag_kin_north}
\end{figure}

\begin{figure}
    \centering
    \includegraphics[clip, trim=0.5cm 0cm 0cm 1cm,width=0.9\linewidth]{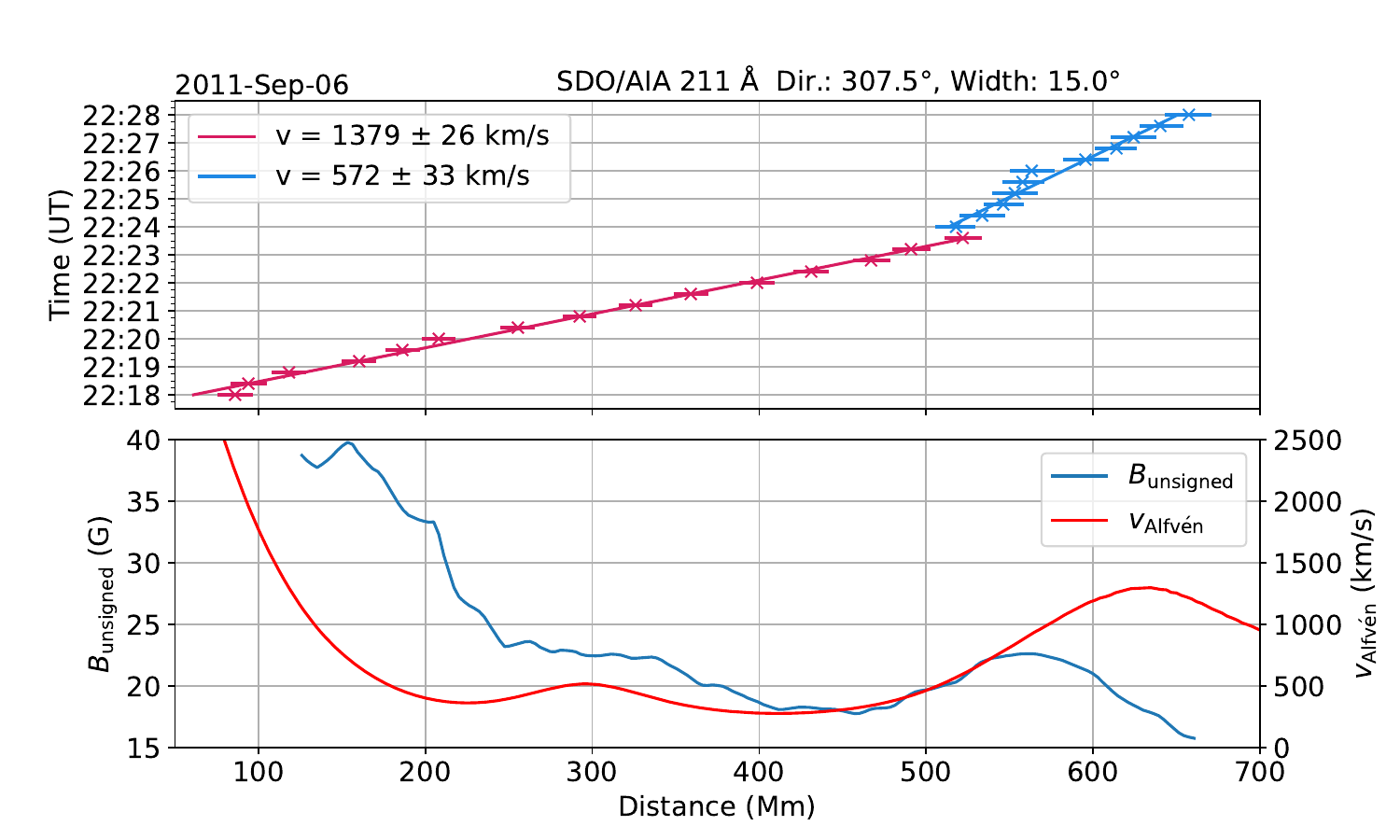}
    \caption{Same as Figure~\ref{fig:mag_kin_north}, but for the sector in the northwest direction.}
    \label{fig:mag_kin_NW}
\end{figure}

Figure~\ref{fig:mag_kin_north} and Figure~\ref{fig:mag_kin_NW} show the resulting magnetic field and Alfv\'{e}n speed profile against distance for the north and northwest direction respectively, together with the kinematics of the wave front. For the north direction (Figure~\ref{fig:mag_kin_north}), the spatial variation in magnetic field indeed seems to coincide with the change in speed at a distance of around $300$~Mm. Further, \citet{vanninathan_plasma_2018} performed a differential emission measurement analysis of the event, showing decreased plasma density towards the North. Both effects combined plausibly explain the change in velocity and the counter-clockwise refraction of the wave in the north direction. For the northwest direction (Figure~\ref{fig:mag_kin_NW}), the wave kinematics do not seem to correlate with variations in the magnetic field distribution or the Alfv\'{e}n speed, except for a distinct reduction in speed at around $550$~Mm coinciding with increased magnetic field strength. This result seems counter intuitive as an increase in field strength results in an increase in magnetosonic speed. Therefore, the decrease in the wave speed observed might be a result of a pile-up of the wave in front of the magnetic obstacle.

Comparing the average Alfv\'{e}n speed in both sectors might give an explanation for the observed velocity difference. Notably, while the  Alfv\'{e}n speed in the north direction drops quickly to, and even below, quiet corona regimes (which is of the order of $150$ to $250$~km~s$^{-1}$ (e.g. \citealt{yuan_measuring_2012,mann_propagation_2023}), the increased magnetic flux towards the Northwest leads to an average Alfv\'{e}n speed of about $500$~km~s$^{-1}$. This is in both directions significantly slower than the speed of the wave, indicating the wave being in the nonlinear regime (which is supported by the high amplitude of the wave of up to $1.37$, \citealt{mann_simple_1995, mann_propagation_2023}) and therefore being influenced only to a certain extent by the surrounding medium. However, the Alfv\'{e}n speed difference is significant enough to justify an effect on the expansion of the different parts of the wave and to explain the speed difference. It should also be noted that the wave travels over a range of heights with changing Alfv\'{e}n speeds, an effect which is not taken into account in more detail in this study. The results show the strength of the SOLERwave tool, using the multi-sector method to create Huygens plots to investigate complex waves with non-isotropic propagation or an unclear wave origin. This can be well paired with the single-sector analysis to investigate the kinematic parameters, like amplitude height over time, of special regions of interest. 




\appendix 
\section{Discussion of the Error Sources}

This appendix describes the different terms of error sources that should be accounted for in the large-scale coronal wave analysis, and is structured as follows: In Section~\ref{appendix:sec_base_ratio}, our choice to use the median of the base ratio values within a segment to represent the segment average as well as the related uncertainty is discussed. Section~\ref{appendix:sec_distance_error} discusses the pixel size error and segment size error, both reducing our certainty in the distance measurement. Note, that we classify terms as ``error" if they are a result of the method used, and uncertainty if they are inherent to the observation. Section~\ref{appendix:sec_direction_error} treats the direction error caused by an offset of the presumed wave origin compared to the center of a circular wave. It also gives an estimation for the distance errors present in the multi-sector method. In Section~\ref{appendix:sec_Projection_Uncertainty} we discuss the projection uncertainty as the result of a three dimensional optically thin structure observed on a two dimensional image plane. Lastly, Section~\ref{appendix:sec_uncertainty_propagation} gives a detailed account of the uncertainty propagation through the single-sector method resulting in the error-bars plotted in the kinematics plots (Figure~\ref{fig:kinematics_northward}, \ref{fig:kinematics_northwest_ward}, \ref{fig:mag_kin_north}, \ref{fig:mag_kin_NW}).



\subsection{Segment Amplitude and attached Uncertainty}
\label{appendix:sec_base_ratio}

\begin{figure}[h!]
    \centering

    \includegraphics[width=0.32\linewidth]{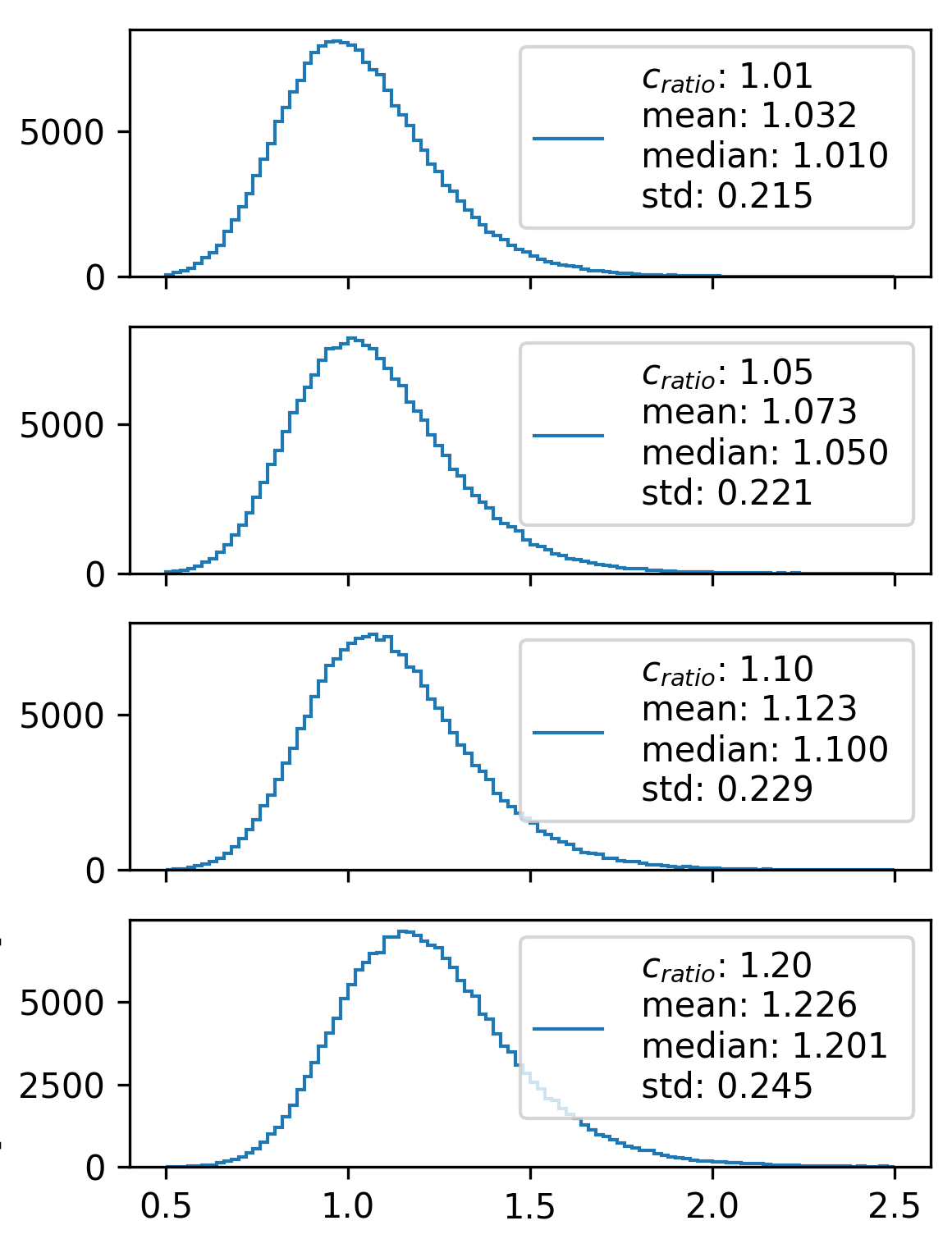}
    \includegraphics[width=0.32\linewidth]{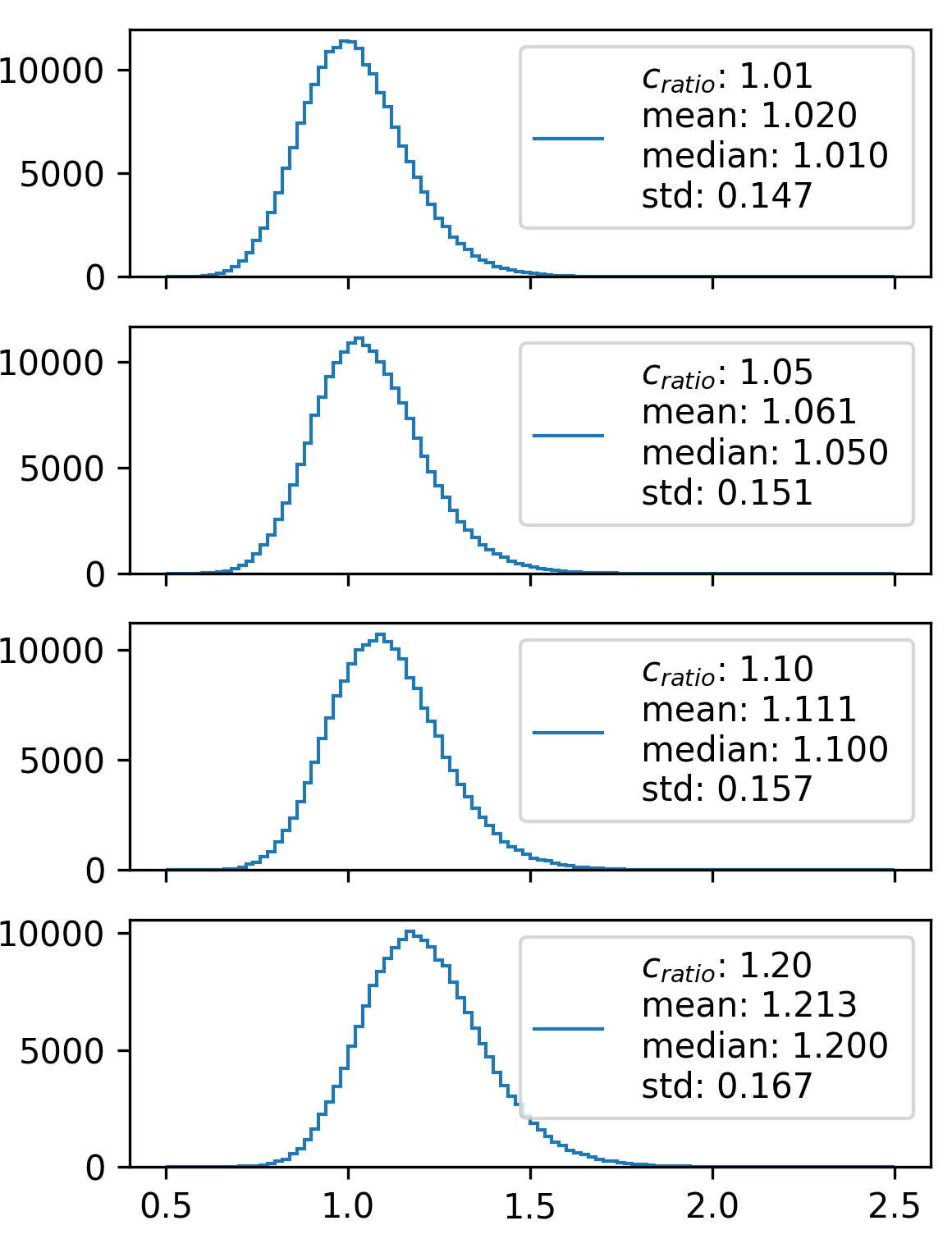}
    \includegraphics[width=0.32\linewidth]{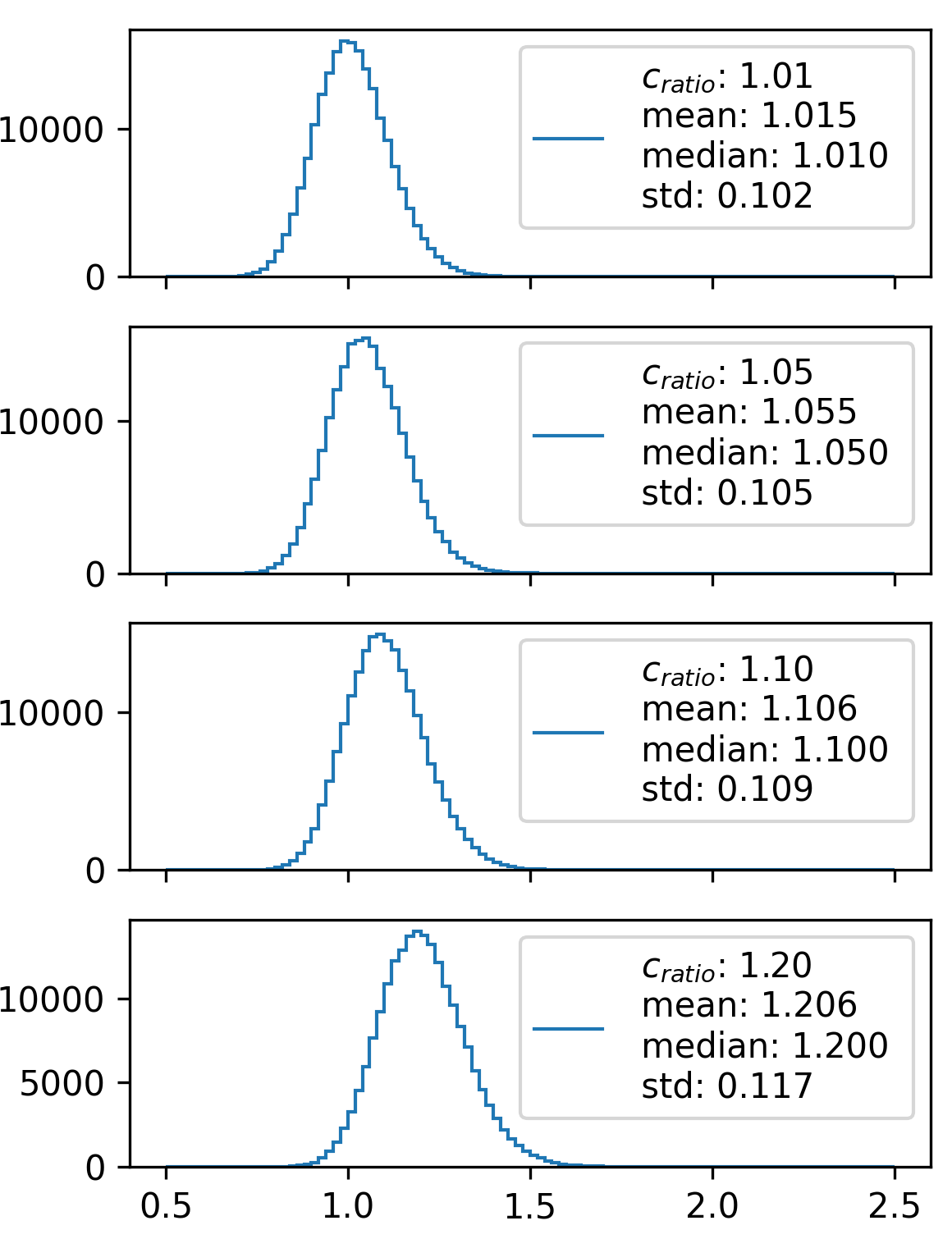}
    \caption{Distributions of the base ratio values, with counts on the $y$-axis and base ratio values on the $x$-axis. For each panel two ensembles with 200000 normal distributed numbers were created. The mean of the first ensemble is given by $\mu = 50, 100, 200$ from the left to the right column, the standard deviation is given by $\sigma = \sqrt{\mu}$. The mean of the second ensemble is given by $\mu' = \mu \cdot c_{\rm ratio}$ with $c_{\rm ratio}$ changing for each row from $1.01$ via $1.05$, $1.1$ to $1.2$, the standard deviation is given by $\sigma' = \sqrt{\mu'}$. For each value of the base ratio ensemble, a value of the second ensemble was divided by a value of the first ensemble. This was repeated for all values of the first and second ensemble, each value was used only once. Thereafter, the median, mean value and standard deviation of the base ratio value distribution were calculated.}
    \label{fig:appendix_base_ratio_distribtuions}
\end{figure}

We investigate the expectation value and its uncertainty of segments derived from base ratio images. The single pixel intensity uncertainty of the images used for the creation of the base ratio image is determined by the physical properties of the measuring instrument (AIA,  \citealt{lemen_atmospheric_2012}) and can be calculated by summing up the different noise terms as shown by \citet{yuan_measuring_2012} using the parameters from \citet{boerner_initial_2012}. They adapted the approach from \citet{aschwanden_time_2000} who used it for the Transition Region and Coronal Explorer (TRACE) mission. For sufficiently large DN values, the shot noise dominates. This leads to a Poisson distribution with the Poisson parameter $\lambda$,  which can be approximated by a Gaussian with the mean $\mu = \lambda$ and standard deviation $\sigma = \sqrt{\lambda}$ for sufficiently large $\lambda \gtrsim 20$ \citep{linden_bayesian_2014}.  Division of these normal distributed values leads to a non-Gaussian distribution, as shown by \citet{marsaglia_ratios_2006}. We investigated the distribution of the base ratio values with a numerical approach. In each test, two ensembles of normal distributed numbers were created: one around a base value in the range $\mu = 50$ to $200$, and one around an elevated value $\mu' = \mu \cdot c_{\rm ratio}$, with $c_{\rm ratio}$ between $1.015$ and $1.3$. The ranges represent pixel values in DN in areas of interest for the SDO/AIA $211~$\AA~passband representative for values prior to the wave event ($\mu$) and during the wave ($\mu'$). As they approximate Poisson distributions,  the standard deviation of the normal distributions is given by the square root of the mean value. The ratio of the numbers between the ensembles were taken and the distribution analyzed. Figure~\ref{fig:appendix_base_ratio_distribtuions} shows a few examples of the created distributions. The results are asymmetric Gaussian-shaped distribution tailed towards larger numbers, especially for small values of $\mu$ and $c_{\rm ratio}$. Calculating both the mean and median of these distributions showed that the median closely followed $c_{\rm ratio}$, while the mean overestimated the value. We therefore use the median of the base ratio values within a segment to represent the segment average, with the standard error of the sample mean as its uncertainty.

\subsection{Distance Errors}
\label{appendix:sec_distance_error}
The distance errors have two main components, the pixel size error and the segment size error. Both emerge from the spherical grid, created by the masks we are using to calculate the perturbation profiles, being finite.

The \textbf{pixel size error} is a result of the segmentation algorithm. To determine the inclusion of a rectangular pixel within a (spherical) segment, the angle of the pixel's center relative to the wave origin is assessed. If the angle is within the segment's threshold, the pixel is included; otherwise, it is not. This results in information within the area of the pixel, but not the area of the segment being included in the averaging process, thereby extending the effective area of the segment. The maximum pixel size error occurs in the direction of the largest gradient (radial from the center) and along the pixels diagonal. For an observation with a $R_\odot \approx 800 ~{\rm pixels}$, equivalent to an SDO/AIA image binned to a size of $2048$ x $2048$ pixels, the pixel size error rises from $0.62$~Mm in the center to above $1.82$~Mm for pixels more than $70$° from the solar center. To determine the correct pixel size error for each segment, the SOLERwave tool searches for the pixel with the largest gradient within each segment and calculates the length along half its diagonal, assuming it would be aligned with the largest gradient.



\begin{figure}[h!]
    \centering
    \includegraphics[clip, trim=5.5cm 13cm 5.5cm 11cm,width=0.5\linewidth]{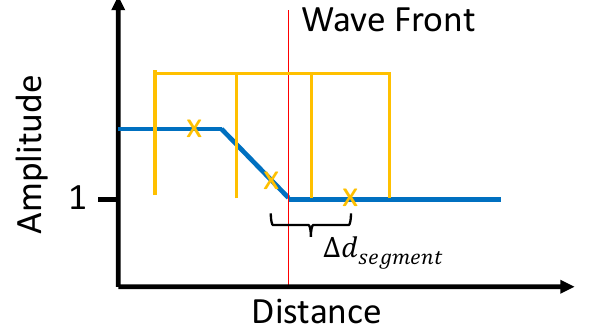}
    \caption{One dimensional sketch of the effects of the distance between segment centers. The blue curve represents the original wave front. Applying a running median filter to it would reproduce the exact curve, as it is monotonically falling. Applying the filter only in certain steps, represented by the orange boxes, results in only the values marked with orange crosses. The supremum of the distance between the real and measured wave front is therefore the distance between segment centers.}
    \label{fig:appendix_wavefront_scetch}
\end{figure}

The \textbf{segment size error} arises from the size of a segment and the finite step width between segments. Figure~\ref{fig:appendix_wavefront_scetch} shows a sketch of the problem at hand, with the falling flank of a wave front (in blue) being sampled with a finite box size (orange). Taking the running median of the monotonically falling parts of a wave front reproduces the sampled curve. This is because, by definition, each value in a monotonically falling function is by definition less than or equal to its predecessor. Therefore, the median of a box derived from such a function is equivalent to the central value, reproducing the original function. The distance error in this case is therefore given by the distance between sampling points ($\Delta d_{\rm segment}$), as the true front could fall anywhere between them.

This analysis only holds true if the underlying function itself has no uncertainty. As shown in Appendix~\ref{appendix:sec_base_ratio}, this is not the case. This leads to a correlation of the segment amplitude uncertainty and the segment distance uncertainty, which is beyond this paper to characterize. Nevertheless, the upper bound for this error can be set assuming that any amount of the wave front within a segment (box) already changes the median. This increases the segment uncertainty $\sigma _{d_{\rm segment}}$ from the distance between two segments, $\Delta d_{\rm segment}$, by adding half the segment width to
\begin{equation}
    \sigma _{d_{\rm segment}} = \Delta d_{\rm segment} + \frac{1}{2} d_{\rm segment}.
\end{equation}

\subsection{Direction Error}
\label{appendix:sec_direction_error}

\begin{figure}[h]
    \centering
    \includegraphics[clip, trim=0.5cm 10cm 1.5cm 8cm,width=0.6\linewidth]{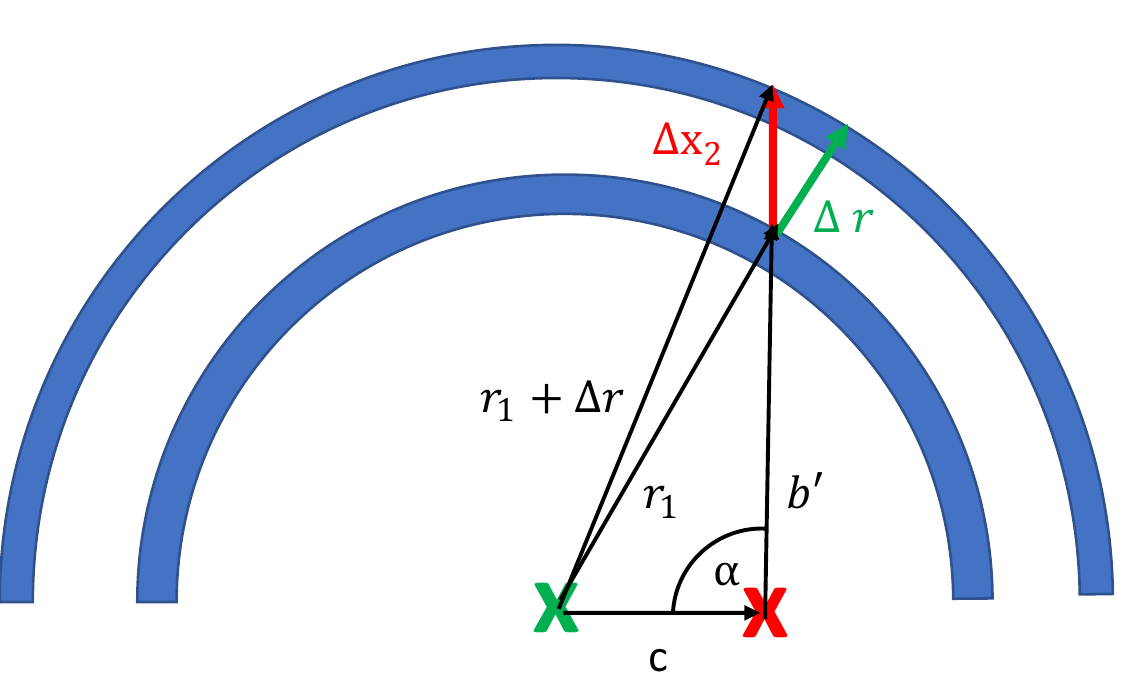}
    \caption{Two dimensional representation of the schematics for the direction error depending on the wave origin. The inner blue semi circle represents the wave front at time $t_{i}$, the outer blue semi circle at $t_{i+1}$. The green cross is the position of the true center of the circular wave, the red cross the position of the shifted center from which the velocity is estimated. All Latin letters indicate angular distances on the solar surface following great circles, the Greek letters annotate angles between those lines.}
    \label{fig:apendix_Directional_Uncertainty}
\end{figure}

The single-sector method assumes a perfectly isotropic propagation causing a circular wavefront on the solar sphere around the presumed wave origin for its derivation of the wave distance and speed. As can be seen, e.g. in Figure~\ref{fig:Huygensplot_Velocities}, this is clearly not the case. Using triangular arithmetics on a sphere (a two-dimensional visual representation of which can be seen in Figure~\ref{fig:apendix_Directional_Uncertainty}) gives for the maximum error
\begin{equation}
    \Delta d_{\rm direction}  = |\Delta r - \Delta x_2|,
\end{equation}
with $\Delta r$ being the radial difference the wave traveled between two points in time and $\Delta x_2$ the perceived distance observed in the case of the wave origin being assumed at the distance $c$ from the true wave origin. The maximum for $\Delta d_{\rm direction}$ is found in the case of $\alpha = 90$°, $\alpha$ being the angle between the observation direction $b'$ and the shift $c$ of the wave center. The exact value of $\Delta d_{\rm direction}$ depends on the distances $c$, $r_1$ and $\Delta r$ involved. For example, for a distance from the true center of the wave $c= 10$° (as measured on the solar surface), a wave that has traveled $r_1 = 20$°$ \approx 240$~Mm with a speed of $1000$~km~s$^{-1}$ (equivalent to a $\Delta r \approx 2$° for $24$~s between observations) would be measured with a speed of  $1140$~km~s$^{-1}$. A reduction in distance to the true center of the wave ($c = 5$° would yield only $1030$~km~s$^{-1}$) reduces this error already significantly.

The multi-sector (Huygens) plotting can somewhat circumvent this problem by deriving the minimum distance between wave front points in subsequent images, and therefore adjusting for the ``true" propagation direction in each element of the wavefront. As the wave fronts are still calculated with the spherical grid, this method has its limits when waves get refracted significantly from the radial (along the surface of the sun) direction. Two main sources of uncertainty arise: First, with increasing azimuthal direction of the wave front, the distance between the intersections of the wave front and the centerline of each sector (the point at which we measure the wave front) increases. This leads to an increase in distance uncertainty, as the true closest points between a wave front at time $t_i$ and $t_{i+1}$ will nearly always fall on the connecting line between two measured positions, not the measurement positions themselves. Second, as for the direction along the great arcs (Appendix~\ref{appendix:sec_distance_error}), the size of the segment results in a segment distance error. As the segment width (corresponding to the sector width, $d_{\rm sector}$) is significantly larger than the segment length $d_{\rm segment}$ ($10$° compared to $1$° in the case of this study), the associated uncertainty is larger. An upper bound for the combined uncertainty can be given by 
\begin{equation}
    \max(\Delta d_{\rm multi-sector}) \approx \sigma _{d_{\rm segment}} \cos(\alpha_{\rm direction}) + \sigma _{d_{\rm sector}} \sin(\alpha_{\rm direction}), 
\end{equation}
with $\alpha_{\rm direction}$ the angle of the velocity vector with respect to the radial great arc passing through the presumed wave origin.

\subsection{Projection Uncertainty}
\label{appendix:sec_Projection_Uncertainty}

\begin{figure}
    \centering
    \includegraphics[clip, trim=0.75cm 0.75cm 0.5cm 0.75cm,width=0.49\linewidth]{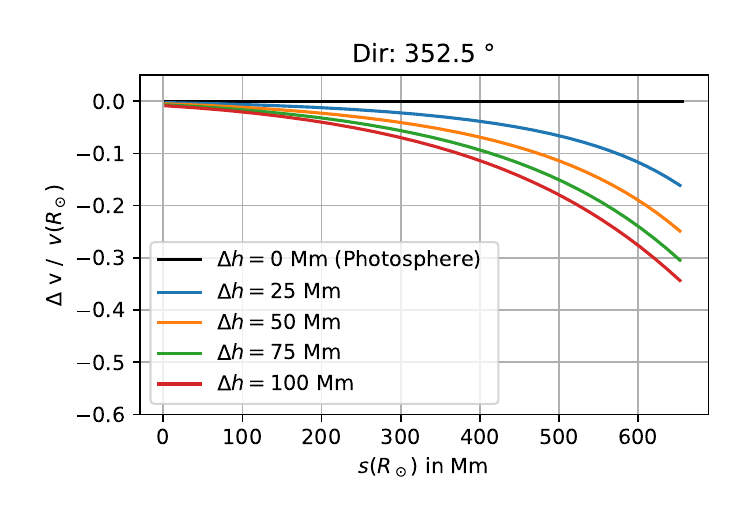}
    \includegraphics[clip, trim=0.75cm 0.75cm 0.5cm 0.75cm,width=0.49\linewidth]{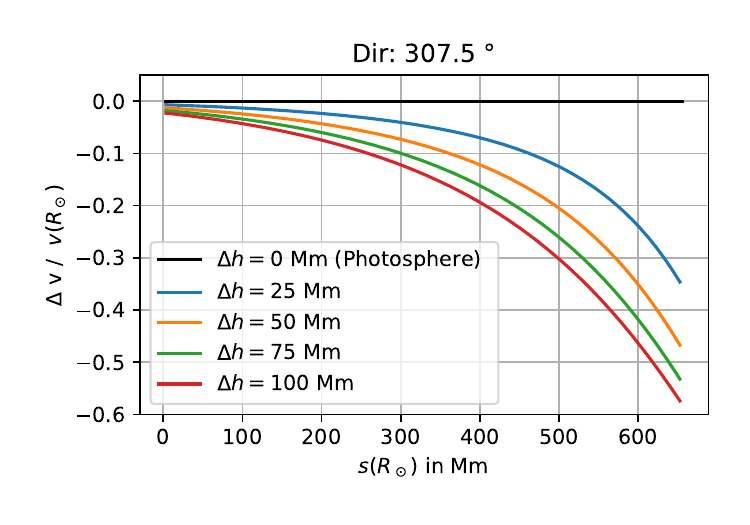}
    \caption{Relative velocity error between parts of the wave assumed to travel at heights $\Delta h$ above the photosphere and parts of wave traveling at photospheric height, i.e. $R_\odot$. The left plot follows the centerline of the northward sector investigated in Section~\ref{sec:sing_sector}, the right plot the centerline of the north-westward sector investigated. The $x$-axis shows distance from the wave origin along these centerlines measured on on photospheric height}
    \label{fig:appendix_Projection_Uncertainty_0deg}
\end{figure}

Large-scale coronal waves are 3D objects, but our observations provide us only with 2D projections onto the image plane. This causes errors due to projection effects. This is further complicated by the emission being from an optically thin plasma, resulting in emission integrated along the line of sight, with contributions from different heights (as shown with simulations by \citealt{hoilijoki_interpreting_2013,downs_validation_2021} and discussed in \citealt{podladchikova_three-dimensional_2019}). Figure~\ref{fig:appendix_Projection_Uncertainty_0deg} shows the relative velocity error of parts of the wave traveling at a height $\Delta h$ above the photosphere (with radius $R_\odot$) of the sectors used in Section~\ref{sec:sing_sector}. The curves were calculated by re-projecting the path $s(R_\odot)$ along a sphere with photospheric radius as observed in SDO/AIA observations (2D images) onto spheres of different heights above the photosphere, i.e. $h = R_\odot + \Delta h$. From the incremental distance between points on the re-projected path ($\Delta s(R_\odot+\Delta h)$), the corresponding incremental distance $\Delta s(R_\odot)$ as measured on the photosphere is subtracted and the result normalized according to 
\begin{equation}
    \frac{\Delta v}{v(R_\odot)} =  \frac{v(R_\odot+\Delta h)-v(R_\odot)}{v(R_\odot)} = \frac{\Delta s(R_\odot+\Delta h)-\Delta s(R_\odot)}{\Delta s(R_\odot)},
\end{equation}
resulting in the relative velocity error by translating the distance increments ($\Delta s$) to the velocities ($v$). The result is an increase in absolute relative velocity error towards the limb of the Sun, as is expected. For a distance of less than $500$~Mm and the bulk of the emission observed coming from coronal heights below about $50$~Mm, the relative velocity error decreases to a maximum of $-10\%$ for the northward direction and $-20\%$ for the north-westward direction. Parts of the wave traveling at about $50$~Mm are therefore in reality $10\%$ respectively $20\%$ slower than when the speed is calculated at photospheric height.

\subsection{General Uncertainty Propagation}
\label{appendix:sec_uncertainty_propagation}

The SOLERwave tool uses a basic uncertainty propagation. The amplitude uncertainty in each segment is estimated as the standard error of the sample mean of the base ratio values within the segment. This is used as relative input weights for the Gaussian fit of the peaks in the perturbation profiles. The uncertainty of the parameters of the Gaussian fit is estimated with the square root of the diagonal elements of the covariance matrix. Further, from the amplitude uncertainty the distance uncertainty between nodes in the perturbation profile is estimated as a linear fit between the amplitude uncertainties of the nodes. As this uncertainty is only used for the front and trailing edge of the peak, we limited it to a maximum of one segment length ($d_{\rm segment}$). 

The segment and pixel error is added symmetrically to the distance uncertainty of the fitted peak and the front/trailing edge. The resulting total distance uncertainties are used as relative weights for the linear fit of the time-distance data. The uncertainty of the wave speed is again estimated by the square root of the diagonal elements of the covariance matrix of the time-distance linear fit. For the uncertainty of the wave width, the distance uncertainties of the front and trailing edges are simply added.

Note that the given uncertainties are often upper bounds. For this reason, the uncertainties were reduced in cases where the upper bound would produce an nonphysical result. The first case to which this applies is the uncertainty of the fitted peak position, which is limited to the range between the wave's front and trailing edges, plus their total distance uncertainties. The second case is the uncertainty of the wave width. For a peak to be detected using the median of segments, at least 50\% of the pixels in a segment must have an amplitude greater than 1. Therefore, the minimum wave width, including the overestimate of the segment error, is kept at half the segment length $d_{\rm segment}$.

%
\begin{acks}

This project has received funding from the European Union's Horizon Europe research and innovation program under grant agreement No 101134999. This article reflects only the authors' views and the European Commission is not responsible for any use that may be made of the information it contains. 
JP acknowledges Research Council of Finland Projects 343581, 364852 and 370793.

\end{acks}

%
%
%
%
%
%
%

%
%
\bibliographystyle{spr-mp-sola}
\bibliography{references}  
%
%
%
%

\end{document}